\documentclass[prl,twocolumn,showpacs,superscriptaddress,amsmath,amssymb]{revtex4-1}

\usepackage{graphicx}% Include figure files
\usepackage{dcolumn}% Align table columns on decimal point
\usepackage{bm}% bold math

\begin{document}

\title{Quantum Griffiths phase inside the ferromagnetic phase of Ni$_{1-x}$V$_x$}

\author{Ruizhe Wang}
\author{Adane Gebretsadik}
\author{Sara Ubaid-Kassis}
\altaffiliation[Present address: ]{Sonoma State University, Department of Engineering Science, Rohnert Park, CA 94928}%Lines break automatically or can be forced with \\
\author{Almut Schroeder}%
%\email{aschroe2@kent.edu}
\affiliation{Physics Department, Kent State University, Kent OH 44242, USA}
\author{Thomas Vojta}
\affiliation{Department of Physics, Missouri University of Science and Technology, Rolla, MO 65409,USA}
\author{Peter J. Baker}
\author{Francis L. Pratt}
\affiliation{ISIS Facility, STFC Rutherford Appleton Laboratory, Harwell Oxford, Oxfordshire, OX11 0QX, UK}
\author{Stephen J. Blundell}
\author{Tom Lancaster}
\altaffiliation[Present address: ]{Durham University, Centre for Materials Physics, South Road, Durham DH1 3LE, UK}%Lines break automatically or can be forced with \\
\author{Isabel Franke}
\author{Johannes S. M\"oller}
\altaffiliation[Present address: ]{Neutron Scattering and Magnetism, Laboratory for Solid State Physics, ETH Z\"{u}rich, CH-8093 Z\"{u}rich, CH    }%Lines break automatically or can be forced with \\
\affiliation{Clarendon Laboratory, Department of Physics, Oxford University, Parks Road, Oxford OX1 3PU, UK}
\author{Katharine Page}
\affiliation{Spallation Neutron Source, Oak Ridge National Laboratory, Oak Ridge, TN}

\date{\today}% It is always \today, today,
             %  but any date may be explicitly specified

\begin{abstract}
We study by means of bulk and local probes the d-metal alloy Ni$_{1-x}$V$_x$ close to the quantum critical concentration, $x_c \approx 11.6\%$, where the ferromagnetic transition temperature vanishes. The magnetization-field curve in the ferromagnetic phase takes an anomalous power-law form with a nonuniversal exponent that is strongly $x$-dependent and mirrors the behavior in the paramagnetic phase.
Muon spin rotation experiments demonstrate inhomogeneous magnetic order and indicate the presence of dynamic fluctuating magnetic clusters. These results provide strong evidence for a quantum Griffiths phase on the ferromagnetic side of the quantum phase transition.

\end{abstract}

%\pacs{71.27.+a, 75.30.Kz, 76.75.+i}
% PACS, the Physics and Astronomy
                             % Classification Scheme.
%\keywords{Suggested keywords}%Use showkeys class option if keyword
                              %display desired
\maketitle

%%%%%%%%%%%%%%%%%%%%%%%%%%%%%%%%%%%%%%%%%%%%%%%%%%%%%%%%%%%%%%%%%%%%%%%%%%%%%%%%%%%%%%%%%%%%%%%%%%%%%
% Introduction
%%%%%%%%%%%%%%%%%%%%%%%%%%%%%%%%%%%%%%%%%%%%%%%%%%%%%%%%%%%%%%%%%%%%%%%%%%%%%%%%%%%%%%%%%%%%%%%%%%%%%

Quantum phase transitions (QPTs) \cite{Sachdev99} continue to be a central topic in condensed matter physics because they are responsible
for a variety of unconventional low-temperature phenomena. For example, the spin fluctuations associated with
QPTs between magnetic and nonmagnetic ground states can lead to non-Fermi liquid behavior or even induce novel phases
of matter \cite{LRVW07}.

Real materials always contain some disorder in the form of vacancies, impurities, and other
defects. In particular, disorder is unavoidable if the QPT is tuned by varying the composition
$x$ in a random alloy such as Ni$_{1-x}$Pd$_x$, CePd$_{1-x}$Rh$_x$, or Sr$_{1-x}$Ca$_x$RuO$_3$.
Research has shown that disorder can dramatically change a QPT and induce
a quantum Griffiths phase, a parameter region close to the transition point that is characterized by
anomalous thermodynamic behavior. This was established for model Hamiltonians
\cite{Fisher92,*Fisher95,ThillHuse95} and later predicted to occur in itinerant magnets
\cite{CastroNetoJones00,VojtaSchmalian05}, superconductors \cite{HoyosKotabageVojta07,*VojtaKotabageHoyos09,DRMS08},
and other systems   (for reviews see, e.g., Refs.\ \cite{Vojta06,*Vojta10}).

Signatures of a magnetic quantum Griffiths phase have been observed, e.g., in diluted Ce compounds \cite{SWKCGG07,*Westerkampetal09}
and, perhaps most convincingly, in the paramagnetic phase of the d-metal alloy Ni$_{1-x}$V$_x$
\cite{UbaidKassisVojtaSchroeder10,SchroederUbaidKassisVojta11}. They consist in anomalous nonuniversal power-law
dependencies of the magnetization, susceptibility and other thermodynamic quantities on temperature and magnetic field
for concentrations $x$ close to but above the quantum critical concentration  $x_c $ (where the ferromagnetic transition
temperature is suppressed to zero). These quantum Griffiths singularities can be attributed to rare magnetic regions embedded
in the paramagnetic bulk, as predicted in the infinite-randomness scenario for disordered itinerant Heisenberg magnets
\cite{VojtaSchmalian05,HoyosKotabageVojta07,*VojtaKotabageHoyos09}.

Do such Griffiths singularities also exist inside the long-range ordered, ferromagnetic phase? Theoretical arguments \cite{SenthilSachdev96,MMHF00} suggest that rare isolated magnetic clusters produce anomalous thermodynamic behavior
on the ferromagnetic side of the QPT as well as on the paramagnetic side. However, the resulting quantum Griffiths singularities
are less universal; depending on the details of the underlying disorder, they range from being stronger than the
paramagnetic ones to being much weaker. So far, clear-cut experimental observations of a quantum Griffiths phase
inside the long-range ordered phase have been missing
\footnote{Unusual scaling behavior in the ferromagnetic phase of URu$_{2-x}$Re$_x$Si$_2$ was initially suggested to
stem from a Griffiths phase but later work showed that this is likely not the case \cite{Baueretal05,ButchMaple09}.}
(see Ref.\ \cite{BrandoBelitzGroscheKirkpatrick16} for a comprehensive
review of QPTs in metallic ferromagnets).
%%%

In this Letter, we
report the results of magnetic measurements and muon spin rotation ($\mu$SR) experiments in
Ni$_{1-x}$V$_x$ across the ferromagnetic QPT. Close to the critical concentration $x_c \approx 11.6\%$,
the dependence of the low-temperature magnetization $M$ on the magnetic field $H$ is well described by anomalous power laws on both
sides of the transition. On the paramagnetic side, $M \sim H^\alpha$ as in earlier work \cite{UbaidKassisVojtaSchroeder10,SchroederUbaidKassisVojta11}.
On the ferromagnetic side, we observe $M-M_0 \sim H^\alpha$ where $M_0$ is the spontaneous magnetization.
The exponent $\alpha$ is strongly $x$-dependent (i.e., nonuniversal) and decreases towards zero at $x_c$.
Strikingly, its $x$-dependence is almost symmetric in $x-x_c$.
$\mu$SR measures the local magnetic fields inside the sample
and reveals the microscopic origins of this anomalous behavior. In the ferromagnetic phase we find a broad distribution
of local magnetic fields signifying inhomogeneous magnetic order. $\mu$SR data for samples close to $x_c$ also indicate that
fluctuating magnetic clusters coexist with the long-range ordered bulk. These results provide strong evidence for a quantum Griffiths
phase on the ferromagnetic side of the QPT in Ni$_{1-x}$V$_x$.

%%%%%%%%%%%%%%%%%%%%%%%%%%%%%%%%%%%%%%%%%%%%%%%%%%%%%%%%%%%%%%%%%%%%%%%%%%%%%%%%%%%%%%%%%%%%%%%%%%%%%
% Experimental stuff
%%%%%%%%%%%%%%%%%%%%%%%%%%%%%%%%%%%%%%%%%%%%%%%%%%%%%%%%%%%%%%%%%%%%%%%%%%%%%%%%%%%%%%%%%%%%%%%%%%%%%

Polycrystalline spherical samples of Ni$_{1-x}$V$_x$ with $x=0$ to 15\% were prepared
and characterized as described in Refs.\ \cite{UbaidKassisVojtaSchroeder10,Schroederetal14}.
A pair distribution function analysis supports the random distribution of the V atoms. Details
of the sample preparation, the characterization with neutron scattering, as well as the magnetization and $\mu$SR measurements
%(using the instruments DOLLY at S$\mu$S and MuSR at ISIS)
(performed at PSI and ISIS)
are summarized in the Supplemental Material
\cite{Supplementary}.

%%%%%%%%%%%%%%%%%%%%%%%%%%%%%%%%%%%%%%%%%%%%%%%%%%%%%%%%%%%%%%%%%%%%%%%%%%%%%%%%%%%%%%%%%%%%%%%%%%%%%
% Phase diagram
%%%%%%%%%%%%%%%%%%%%%%%%%%%%%%%%%%%%%%%%%%%%%%%%%%%%%%%%%%%%%%%%%%%%%%%%%%%%%%%%%%%%%%%%%%%%%%%%%%%%%

\begin{figure}
\includegraphics[width=\columnwidth]{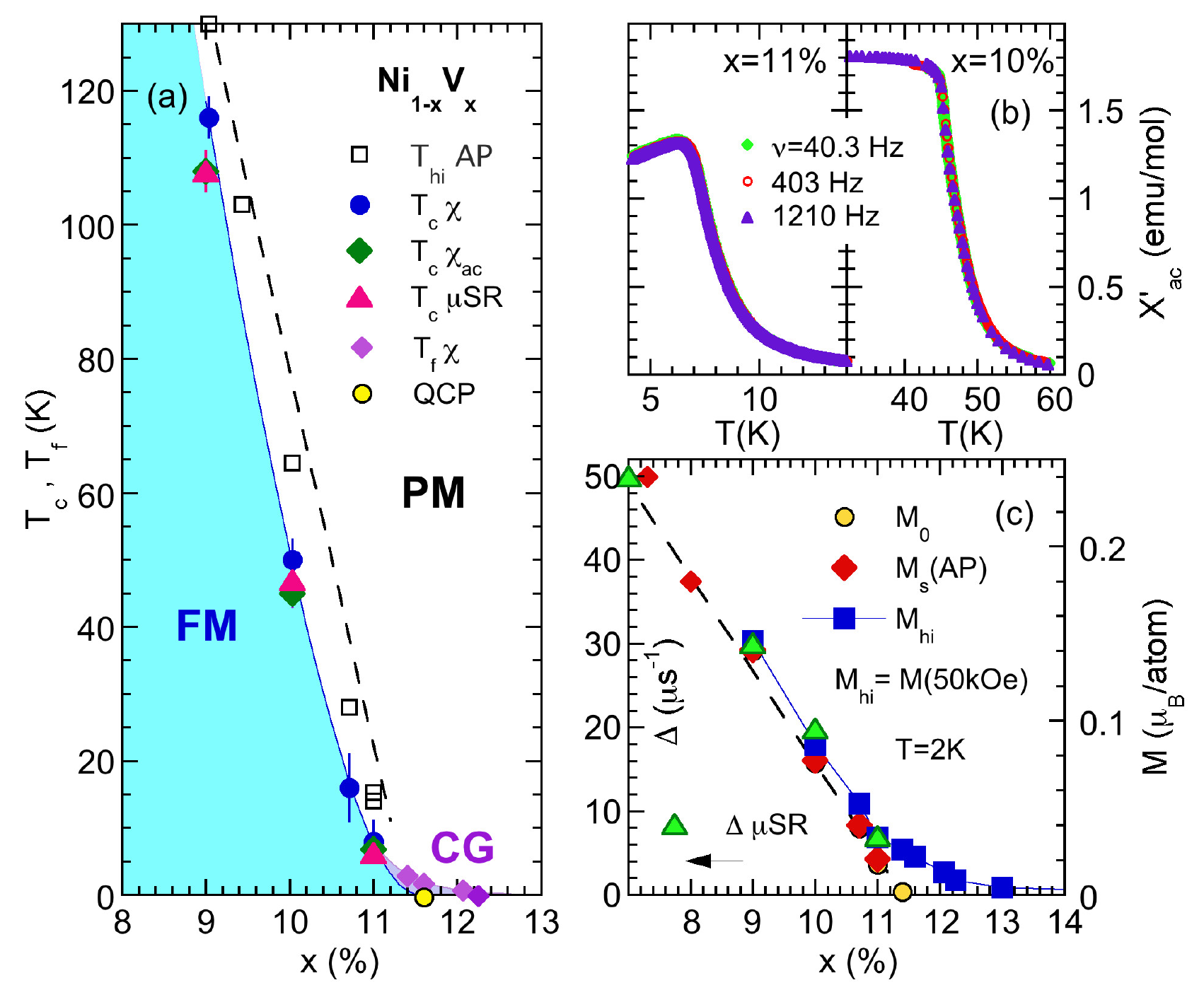}% Fig1 for intro overview
\caption{(a) Phase diagram of Ni$_{1-x}$V$_x$ showing paramagnetic (PM), ferromagnetic (FM), and cluster glass (CG) phases. The ferromagnetic transition
temperature $T_c$ is found using three different methods (see text), leading to a quantum critical point (QCP) at $x_c=11.6\%$.
The high-field (Arrott plot) estimate $T_{hi}$ of the transition shows a linear $x$-dependence (dashed line).
(b) ac-susceptibility $\chi'_{ac}$ vs. temperature $T$ (absolute scale estimated by dc-M). $T_c$ is marked by a cusp independent of frequency $\nu$
%($(\delta T/T_c)/log\nu<0.005/decade$).
(c) Zero-field moment $M_s$ (from Arrott plots), $M_0$ (from $M(H)$ power law) and $\mu$SR field distribution width $\Delta$ show linear $x$-dependencies (dashed line).
$M_{hi}$ is the magnetization in a field of $H=50$\,kOe. Data of $T_{hi}$ and $M_s$ from \cite{boelling68,gregory75C} included.}
\label{fig:phase_diagram}
\end{figure}

At first glance, Ni$_{1-x}$V$_{x}$ features a simple phase diagram: The ferromagnetic ordering temperature $T_c$ and the spontaneous magnetization
$M_s$ are linearly suppressed with increasing $x$ and vanish between $x=11\%$ and 12\%,  as shown in Figs.\ \ref{fig:phase_diagram}(a)
and \ref{fig:phase_diagram}(c).
This critical concentration is much smaller than the corresponding $x_c=97.5\%$ for Ni$_{1-x}$Pd$_x$ \cite{NBKMTL99} because the V atoms, with 5 fewer d-electrons than Ni,
also suppress the spins of their Ni neighbors and thus create large defects \cite{Friedel58,CollinsLow65}.
%rather than just replacing single magnetic Ni sites with ``non magnetic" V atoms.
The inhomogeneous suppression of magnetic order causes deviations from the linear $x$-dependence of $T_c$  close to the critical concentration.
We determined $T_c$ from the maximum of the susceptibility $dM/dH(T,H\rightarrow0)$ \cite{SchroederUbaidKassisVojta11}, the cusp in the ac susceptibility
$\chi'_{ac}(T,H=0)$ \cite{Wangetal15} (see Fig.\ \ref{fig:phase_diagram}(b)), and the onset of the zero-field $\mu$SR amplitude $A_{FM}(T)$ \cite{Schroederetal14}
(see Fig.\ \ref{fig:muon} below).
All estimates agree well with each other. The resulting $T_c(x)$ curve develops a tail and follows the prediction \cite{HoyosKotabageVojta07,*VojtaKotabageHoyos09}
of the infinite-randomness scenario, giving a critical concentration $x_c=11.6\%$. (In contrast, the tail is absent when an ordering temperature $T_{hi}$ is estimated
via extrapolation from high fields, e.g, via standard Arrott plots of $H/M$ vs.\ $M^2$.)
%(A tail of the x-dependence of the ordering temperature $T_c$ or $T_f$ and the change from long range to short range order has been reported in other diluted systems \cite{westerkamp09CePd,pikul12,bauer05}.)

The actual quantum critical point at $T=0$ and $x=x_c$ is masked by a cluster glass phase
that appears for $x\gtrsim 11.4\%$ below a freezing temperature $T_f \leq 3$\,K, see Fig.\ \ref{fig:phase_diagram}(a)
\cite{UbaidKassisVojtaSchroeder10,Wangetal15}.
It is rapidly suppressed by small dc fields and does not affect the physics considered in this Letter.

%%%%%%%%%%%%%%%%%%%%%%%%%%%%%%%%%%%%%%%%%%%%%%%%%%%%%%%%%%%%%%%%%%%%%%%%%%%%%%%%%%%%%%%%%%%%%%%%%%%%%
% M vs. H
%%%%%%%%%%%%%%%%%%%%%%%%%%%%%%%%%%%%%%%%%%%%%%%%%%%%%%%%%%%%%%%%%%%%%%%%%%%%%%%%%%%%%%%%%%%%%%%%%%%%%

\begin{figure}
\includegraphics[width=\columnwidth]{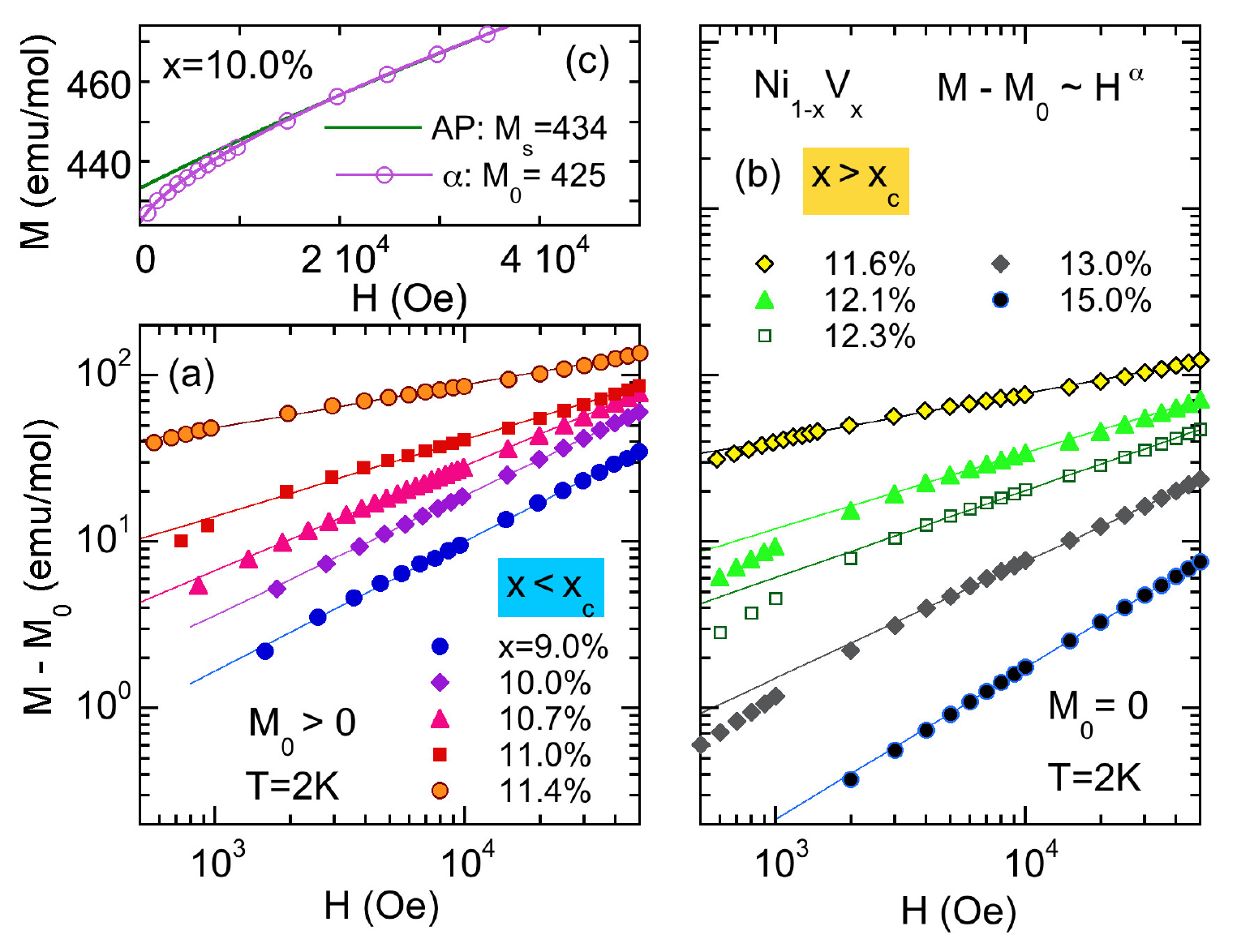}% Fig2 alpha
\caption{Magnetization $M$ vs (internal) magnetic field $H$ for several compositions $x$ at the lowest $T=2$\,K. An offset $M_0(x) >0 $ has been subtracted for $x<x_c$ in (a); $M_0=0$
for $x\ge x_c$ in (b). Solid lines represent fits to $M(H) = M_0 + d_\alpha H^\alpha$. (c) $M$ vs $H$ for $x=10\%$ with power-law fit defining $M_0$ and Arrott plot (AP) fit defining $M_s$.}
\label{fig:MvsH}
\end{figure}

We now analyze the field dependence of the magnetization $M$ at low $T$.
%on the magnetic field $H$.
Figure \ref{fig:MvsH} shows $M$ vs.\ $H$
at $T=2$\,K for V concentrations $x$ on both sides of the QPT.
For paramagnetic samples ($x \ge x_c=11.6\%$), the magnetization follows the anomalous power law $M(H) =d_\alpha H^\alpha$ over an extended field
range from about 2\,kOe to the highest available field of 50\,kOe. Interestingly,
the field dependence of the magnetization  in the long-range ordered ferromagnetic phase ($x<x_c$) is also
well described by a power-law form, viz., $M(H) = M_0 + d_\alpha H^\alpha$ where $M_0$ represents the nonzero spontaneous magnetization.
As in the paramagnetic phase, these power laws hold in a wide field range from about 1 or 2\, kOe to 50\,kOe
(while the conventional Arrott plot description breaks down below about 10\,kOe, see Fig.\  \ref{fig:MvsH}(c)).

\begin{figure}
\includegraphics[width=\columnwidth]{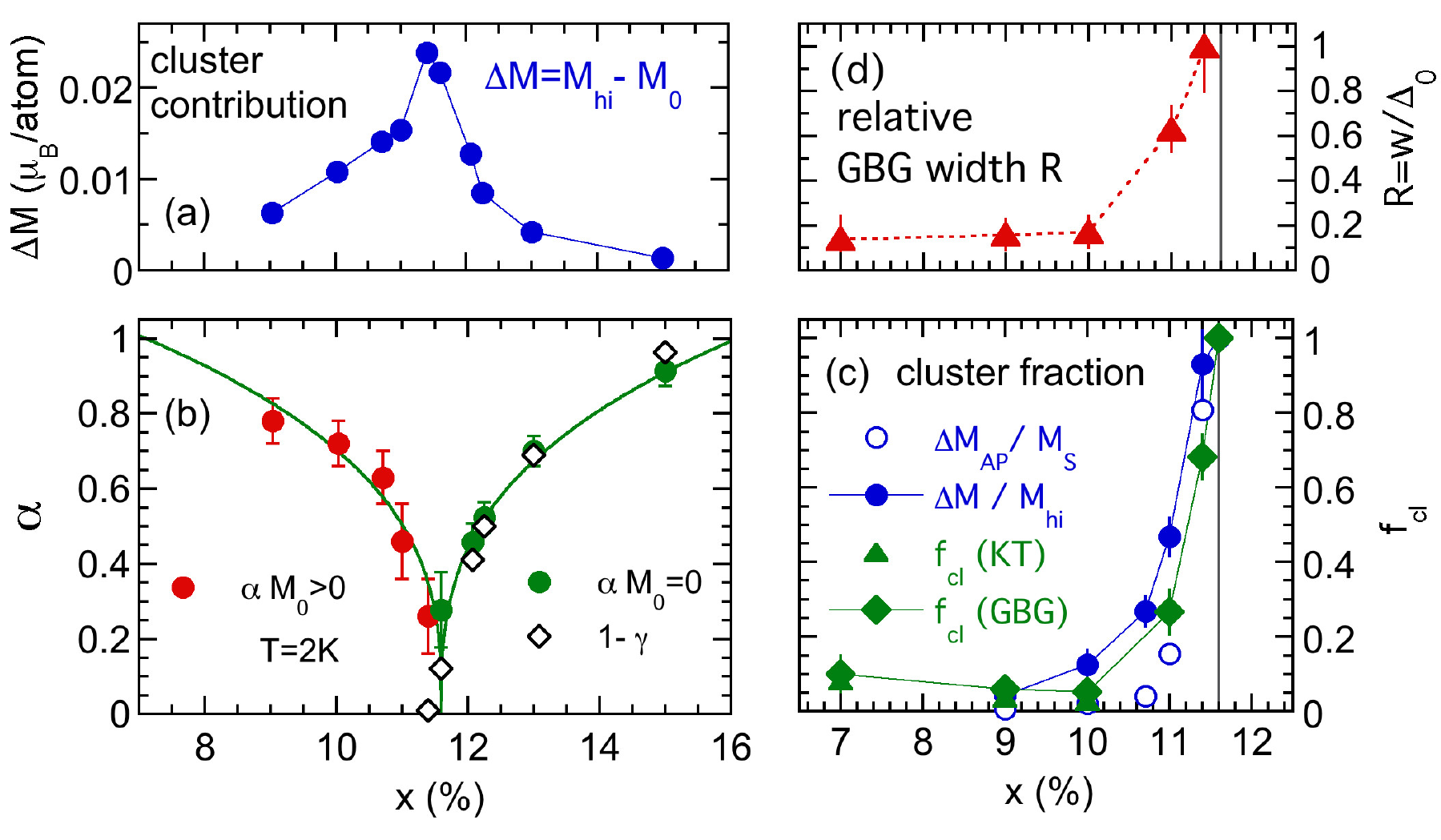}% Fig3
\caption{(a) Cluster contribution $\Delta M = M_{hi}-M_0$ with $M_{hi}=M$(50kOe) vs.\ concentration $x$ in Ni$_{1-x}$V$_x$.
(b) Non-universal exponent $\alpha$ vs.\ $x$, and susceptibility exponents  $\gamma$ from Ref.\ \cite{UbaidKassisVojtaSchroeder10}.
Lines  are  universal power-law fits $\alpha(x)\sim|x-x_c|^{\nu\psi}$. (c) Cluster fraction $f_{cl}$ vs.\ $x$ from different methods.
(d) Relative width $R=w/\Delta_c$ vs.\ $x$ of the Gaussian-broadened Gaussian used in the $\mu$SR analysis. Data evaluated at lowest temperature ($1.5-2$\,K for $x>10\%$).}
\label{fig:clusterratio}
\end{figure}

The exponent $\alpha$ is nonuniversal, i.e., strongly $x$-dependent. It has a minimum close to the critical concentration $x_c$
and increases monotonically towards the linear-response value $\alpha=1$ with increasing distance
from $x_c$, as shown in Fig.\ \ref{fig:clusterratio}(b).
Strikingly, the $\alpha(x)$ curve is nearly symmetric in $x-x_c$. It can be fitted with a power law,
$\alpha(x) \sim |x-x_c|^{\nu\psi}$
%giving the
with exponent $\nu\psi\approx 0.34 \pm 0.08$ \footnote{This estimate of $\nu\psi$ is somewhat smaller than the value $\nu\psi=0.42$ obtained in Ref.\ \cite{UbaidKassisVojtaSchroeder10}
from the $x$-dependence of the susceptibility exponent $\gamma$. The deviation is within the error bars.}
confirming $x_c=11.6\% \pm 0.1\%$.
%if fitted separately xc=11.4-11.7, if fitted together 11.6 sharp, so compromise
%if xc=11.6 fixed expo=0.34 for both sides error in expo=0.08 an FM side

What is the origin of these unusual mag\-neti\-zation-field curves? In the paramagnetic phase, they can be attributed to
magnetic clusters that are embedded in the paramagnetic bulk \cite{UbaidKassisVojtaSchroeder10,SchroederUbaidKassisVojta11}.
These clusters exist on rare Ni-rich regions in the sample. Their slow independent fluctuations lead to anomalous power laws,
the Griffiths singularities, in the temperature and field dependencies of various
thermodynamic quantities \cite{Vojta06,*Vojta10}.
Deviations at the lowest fields and temperatures stem from weak interactions between the rare regions
that freeze their dynamics \cite{DobrosavljevicMiranda05,UbaidKassisVojtaSchroeder10}
\footnote{If the relevant rare regions are so large that order parameter conservation hampers the dynamics of the clusters,
the functional form of the Griffiths singularities changes \cite{NozadzeVojta12}. Based on a typical cluster moment
of about 12\,$\mu_B$ at $x_c$ \cite{UbaidKassisVojtaSchroeder10,SchroederUbaidKassisVojta11}, this likely does not play a role here.}.
Our observation of anomalous magnetization-field curves
\emph{below} the critical concentration $x_c$ indicates that
disconnected magnetic clusters that fluctuate independently from the long-range ordered bulk
also play a crucial role inside the ferromagnetic phase.

To analyze the importance of these clusters quantitatively, we estimate their contribution to the magnetization.
A conservative estimate can be obtained by comparing the spontaneous magnetization $M_0$ with the zero-field magnetization
$M_s$ obtained via Arrott plot extrapolation from high fields (see in Fig. \ref{fig:phase_diagram}(c)). As the clusters are disconnected from the bulk, they do not
contribute to $M_0$. In high fields they are fully polarized, however, and thus included in $M_s$. Consequently,
$\Delta M_{AP}= M_s-M_0$ measures the cluster contribution to $M$.
%and $\Delta M_{AP}/M_s$ characterizes the cluster fraction.
Alternatively, one could simply evaluate $\Delta M = M_{hi}-M_0$ with $M_{hi}=M(H=50\,\textrm{kOe})$ and define the cluster fraction as
$\Delta M / M_{hi}$ \footnote{This measure somewhat overestimates the cluster contribution because
$M_{hi}$ in $H=$ 50 kOe also contains the bulk response to the field.}.
The $x$-dependence of $\Delta M$ is shown in Fig.\ \ref{fig:clusterratio}(a). $\Delta M$ has a maximum close to $x_c$
and decreases for $x>x_c$ because the total number of magnetic Ni atoms decreases.
$\Delta M$ also decreases for $x<x_c$ because it becomes less likely that a magnetic cluster
remains disconnected from the bulk. By comparing $\Delta M$ with the typical cluster moment of 12\,$\mu_B$
\cite{UbaidKassisVojtaSchroeder10,SchroederUbaidKassisVojta11}, we estimate a cluster density at $x_c$ of about one cluster
per 500 Ni atoms.
Figure \ref{fig:clusterratio}(c) presents the cluster fractions $\Delta M_{AP}/M_s$ and $\Delta M / M_{hi}$ as functions of $x$.
The  measures track each other and indicate that clusters become relevant for $x > 10\%$.

%%%%%%%%%%%%%%%%%%%%%%%%%%%%%%%%%%%%%%%%%%%%%%%%%%%%%%%%%%%%%%%%%%%%%%%%%%%%%%%%%%%%%%%%%%%%%%%%%%%%%
% muSR
%%%%%%%%%%%%%%%%%%%%%%%%%%%%%%%%%%%%%%%%%%%%%%%%%%%%%%%%%%%%%%%%%%%%%%%%%%%%%%%%%%%%%%%%%%%%%%%%%%%%%

To gain microscopic insight into these clusters and their dynamics, we employ $\mu$SR
experiments (see, e.g., Ref.\ \cite{Blundelleasy} for an introduction and Ref.\ \cite{YaouancReotier11} for a technical
review).
%In this technique, spin-polarized muons are implanted in the sample. Their spins then precess in the
%local magnetic field at the stopping site. After an average life time of 2.2$\mu$s, the muon decays into a
%positron and two neutrinos with the positron emitted preferably in the direction of the muon spin.
%lastversion:
In this technique, spin-polarized positive muons are implanted in the sample. Their spins then precess in the local magnetic field at the stopping site until the muon decays, with a positron emitted preferentially in the direction of the muon spin. Analyzing the
asymmetry $A(t)$ of the positron emission
as a function of time
thus gives direct access to the distribution of \emph{local} magnetic fields in the sample.
$\mu$SR played an important role in characterizing unconventional magnetism, e.g., in heavy-fermion compounds \cite{Amato97}, spin glasses \cite{uemuraSG}, and disordered, non-Fermi liquid metals \cite{McLaughlin04}.
As $\mu$SR experiments are sensitive towards small magnetic moments, spatial inhomogeneities, and slow fluctuations,
they are well suited to identify and study magnetic clusters.

Data for the  muon asymmetry $A(t)$ in zero magnetic field for
several samples
%covering the concentration range
%Ni$_{1-x}$V$_x$
from $x=0$ to $12.3\%$ are presented in the Supplemental Material \cite{Supplementary}, together with further details of the
%muon
analysis. For pure Ni ($x=0$), $A(t)$ features a single (nearly undamped) precession frequency confirming a uniform local magnetic field and thus uniform ferromagnetic order.
In contrast, the $x=12.3\%$ sample on the paramagnetic side of the QPT
shows a very weak depolarization. It can be described by a simple exponential decay,
$A(t)= A_{PM} P_{PM}(t) = A_{PM} \exp(-\lambda t)$,   caused by quasistatic diluted V nuclear spins as well as by fluctuating Ni clusters in the extreme motional narrowing limit.

Here, we focus on two samples ($x=10\%$ and 11\%) that are close to the QPT but on its ferromagnetic side.
At low temperatures,
%the muon asymmetry
$A(t)$ of
the
$x=10\%$
sample
(shown in Fig.\ \ref{fig:muon}(a)) features a single dip but no further oscillations.
\begin{figure}
\includegraphics[width=\columnwidth]{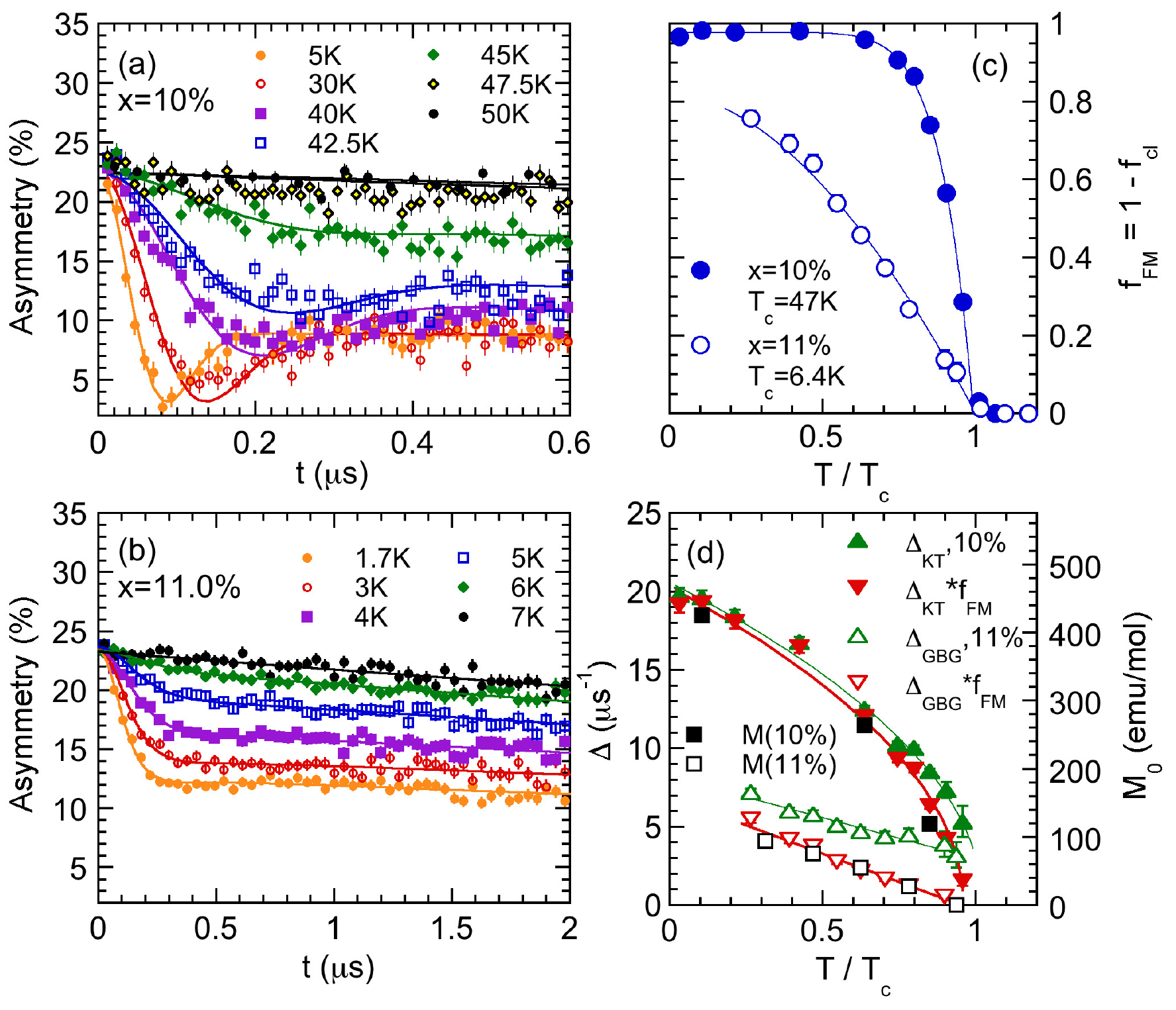}% Fig4 ZFmusR
\caption{(a,b) $\mu$SR asymmetry $A$ vs.\ time $t$ for different concentrations $x$ and temperatures $T$
(collected at DOLLY, S$\mu$S).
Lines represent fits to Eq.\ (\ref{eq:A(t)}) using different
%forms of
$P_{FM}(t)$:
 KT form
 %with distribution $\Delta$
 for $x=10\%$ (a), GBG form
 %with effective Gaussian broadened Gaussian width $\Delta_{GBG}$
 for $x=11\%$ (b),
 (for details see text).
(c) ferromagnetic  fraction (amplitude ratio)  $f_{FM}$ vs.\ temperature $T$. (d) Field distribution width $\Delta=\Delta_{KT}$ for $x=10\%$ and  $\Delta=\Delta_{GBG}$ for $x=11\%$  in frequency units $\Delta=\gamma_{\mu} \langle B^2_{loc}\rangle^{1/2}$ (with $\gamma_{\mu}=2\pi \times135.5$\,MHz/T).
The magnetization $M_0$ and $\Delta$ are proportional to each other
(with $M_0/\Delta \approx 23$\,emu/mol MHz), but only if $\Delta$ is scaled by $f_{FM}$.
 }
\label{fig:muon}
\end{figure}
Analogous behavior is observed for
%all samples with
$7\%\leq x \leq 10\%$ \cite{Schroederetal14}.
It can be described by a Gaussian distribution of local magnetic fields
of width $\Delta$, leading to  $A(t)= A_{FM} P_{FM}(t)$ with $P_{FM}(t)=P_{KT}(t;\Delta) $ where
$P_{KT}(t;\Delta)$ is the well-known Kubo-Toyabe (KT) depolarization function \cite{Hayanoetal79}.
At temperatures below about 0.5 $T_c$, the data follow the static KT form,
signifying a moderately inhomogeneous, long-range ordered state.

Over the entire temperature range, $A(t)$ can be modeled by two components
(with temperature-dependent amplitudes) and a small constant background term,
%(e.g. $A_{BG}=2-3\%$ for similar silver mount in Dolly).
\begin{equation}
A(t)=A_{PM}(T) P_{PM}(t) + A_{FM}(T) P_{FM}(t) + A_{BG}~.
\label{eq:A(t)}
\end{equation}
The temperature dependence of the relative amplitude $f_{FM}=A_{FM}/(A_{FM}+A_{PM})$ which represents the
FM fraction of the sample is presented in Fig.\ \ref{fig:muon}(c). It rapidly increases as the temperature is
lowered below $T_c$ and reaches values close to unity for $T\le  0.7\, T_c$.
The width $\Delta$ of the local magnetic field distribution increases with decreasing $T$; below about
$0.7\, T_c$,
$\Delta \propto M_0$
%,it is proportional to the spontaneous magnetization $M_0$,
as shown in Fig.\
\ref{fig:muon}(d).

For $x=11\%$, the KT form fails to describe $A(t)$ (shown in Fig.\ \ref{fig:muon}(d))
as the typical dip is missing;
data taken in longitudinal fields also exclude a dynamic KT form \cite{Schroederetal14}.
A nearly static broader-than-Gaussian field distribution can account for the main, fast time dependence of $A(t)$. In fact, $A(t)$ can be fitted well using Eq.\ (\ref{eq:A(t)}) with $P_{FM}= P_{GBG}(t;\Delta_0, w) $
where $P_{GBG}$ is the static ``Gaussian-broadened Gaussian" (GBG) depolarization function suggested in Ref.\ \cite{NoakesKalvius97}, and $\Delta_0$ and $w$ are the average and width of the Gaussian
of Gaussians.
The temperature dependencies of the effective distribution width $\Delta_{GBG}=(\Delta_0^2+ w^2)^{1/2}$  and of the relative amplitude $f_{FM}$ are shown in Figs.\ \ref{fig:muon}(d) and (c).
%\footnote{A dynamic form of $P_{GBG}$ does not change the presented parameters, e.g. it yields a negligibly  small $\nu$ for low $T$ (similar to $x=10\%$),
%and towards $T_c$ higher values of $\nu$ that are dependent on $\lambda_{PM}$,
%see Ref.\ \cite{Schroederetal14}.}.
The need for a broad field distribution to describe the ferromagnetic component indicates strongly inhomogeneous order.
Moreover, the ferromagnetic ratio $f_{FM}$
%shown in Fig.\ \ref{fig:ZFparam}(d)
increases only slowly below $T_c$, and a sizable paramagnetic
contribution representing about 20\% of the sample volume remains even at the lowest $T$.  This paramagnetic contribution stems from the fluctuating moments of Ni-rich clusters that are disconnected
from the long-range ordered bulk.

The cluster fraction $f_{cl}=1-f_{FM}$ can be obtained for all $x$ using KT and GBG fits of $A(t)$ at the lowest $T$. As shown in Fig.\ \ref{fig:clusterratio}(c),
these  $\mu$SR based cluster fractions agree well with the
estimates from the magnetization data and indicate that clusters are relevant for $x>10\%$. Accordingly, the relative width
$R=w/\Delta_0$ of the Gaussian of Gaussians \cite{NoakesKalvius97} in the field distribution starts increasing for $x>10\%$, as shown in
Fig.\  \ref{fig:clusterratio}(d).

%%%%%%%%%%%%%%%%%%%%%%%%%%%%%%%%%%%%%%%%%%%%%%%%%%%%%%%%%%%%%%%%%%%%%%%%%%%%%%%%%%%%%%%%%%%%%%%%%%%%%
% Conclusions
%%%%%%%%%%%%%%%%%%%%%%%%%%%%%%%%%%%%%%%%%%%%%%%%%%%%%%%%%%%%%%%%%%%%%%%%%%%%%%%%%%%%%%%%%%%%%%%%%%%%%

In summary, we studied the d-metal alloy Ni$_{1-x}$V$_x$ close to its quantum-critical concentration $x_c$,
focusing on the ferromagnetic side of the QPT.
We found that the low-temperature magnetization-field curve in the ferromagnetic phase
follows the power-law  $M(H)=M_0 +d_\alpha H^\alpha$ in analogy to the power-law
Griffiths singularity $M(H) \sim H^\alpha$ on the paramagnetic side. This anomalous behavior can
be attributed to
%the contributions from
magnetic clusters existing on
disconnected rare Ni-rich regions of the sample. Further evidence for such clusters
%came
comes from $\mu$SR experiments that
%revealed
reveal strongly inhomogeneous magnetic order and the presence
of paramagnetic, fluctuating moments inside the long-range ordered ferromagnet (for samples sufficiently
close to $x_c$).
These results provide evidence for a quantum Griffiths phase inside the ferromagnetic phase and demonstrate that QPTs in strongly disordered systems are qualitatively different not just from their clean counterparts but also from disordered classical phase transitions. Disorder at a classical transition may change its universality class or turn a first-order transition continuous. In contrast, we  observed much stronger effects. Thermodynamic and other properties of Ni$_{1-x}$V$_x$ close to its QPT are dominated by rare events, resulting, for example, in a diverging magnetic susceptibility not just at $x_c$ but over a range of $x$
close to $x_c$.

In theoretical studies of model Hamiltonians \cite{SenthilSachdev96,MMHF00}, quantum Griffiths phases on the magnetic side of the QPT are much less universal than those on the paramagnetic side. This stems from the fact that the probability of finding a magnetic cluster that is disconnected from the long-range ordered bulk of the system depends on the
details of the disorder.
Specifically, in a percolation scenario, a magnetic cluster can be isolated by a surface (shell) of nonmagnetic sites (or broken bonds). Such events have a comparatively high probability; the resulting Griffiths singularities on the ferromagnetic side are thus expected to be stronger than power laws, i.e., stronger than their paramagnetic analogs \cite{SenthilSachdev96}.
For weak disorder, in contrast, a cluster has be far away from the long-range ordered bulk to be isolated.
This reduces the cluster probability and leads to ferromagnetic Griffiths singularities that are weaker than the power laws on the paramagnetic side \cite{MMHF00}.
The disorder in Ni$_{1-x}$V$_x$ is not purely percolational because the material is a metal,
but it is rather strong because each V atom creates a large local defect. The strength of the quantum Griffiths singularities is therefore expected to be between the above limiting cases, in agreement with our observations. However,
the existing theories cannot explain the striking symmetry in $x-x_c$ of the Griffiths singularities
found here \footnote{Such a symmetry does occur in certain one-dimensional random spin chains \cite{Fisher92,*Fisher95},
but it does not generalize to higher dimensions.}. This remains a challenge for future work.

%copyright statement%
%This manuscript has been co-authored by UT-Battelle, LLC under Contract No. DE-AC05-00OR22725 with the U.S. Department of Energy. The United States Government retains and the publisher, by accepting the article for publication, acknowledges that the United States Government retains a non-exclusive, paid-up, irrevocable, world-wide license to publish or reproduce the published form of this manuscript, or allow others to do so, for United States Government purposes. The Department of Energy will provide public access to these results of federally sponsored research in accordance with the DOE Public Access Plan $(http://energy.gov/downloads/doe-public-access-plan)$.

%%%%%%%%%%%%%%%%%%%%%%%%%%%%%%%%%%%%%%%%%%%%%%%%%%%%%%%%%%%%%%%%%%%%%%%%%%%%%%%%%
%\section*{Acknowledgements}
%%%%%%%%%%%%%%%%%%%%%%%%%%%%%%%%%%%%%%%%%%%%%%%%%%%%%%%%%%%%%%%%%%%%%%%%%%%%%%%%%

This work was supported by the NSF under Grant No.\ DMR-1506152 and by EPSRC (UK). We are grateful for the provision of beam time at the STFC ISIS Muon Facility and S$\mu$S, Paul Scherrer Institut, Switzerland. This work has benefitted from the use of NPDF at the Los Alamos Neutron Science Center, Los Alamos National Laboratory, funded by the US Department of Energy.  Part of this research was conducted at the NOMAD instrument at the Spallation Neutron Source, a US Department of Energy Office of Science User Facility operated by Oak Ridge National Laboratory.

\bibliographystyle{apsrev4-1}
%\bibliography{../00Bibtex/rareregions}
%\bibliography{citations_tv_5mod}
%

%%%%%%%%%%%%%%%%%%%%%%%%%%%%%%%%%%%%%%%%%%%%%%%%%%%%%%%%%%%%%%%%%%%%%%%%%%%%%%%%%%%%%%%%%%%%%%%%%%%%%%%%
\clearpage
%%%%%%%%%%%%%%%%%%%%%%%%%%%%%%%%%%%%%%%%%%%%%%%%%%%%%%%%%%%%%%%%%%%%%%%%%%%%%%%%%%%%%%%%%%%%%%%%%%%%%%%%
% merge with supplemental material

\onecolumngrid
\begin{center}
{\large\bf Supplemental material for\\
Quantum Griffiths phase inside the ferromagnetic phase of Ni$_{1-x}$V$_x$}
\end{center}

\bigskip
\twocolumngrid

\setcounter{equation}{0}
\setcounter{figure}{0}
\setcounter{table}{0}
\setcounter{page}{1}
\makeatletter
\renewcommand{\theequation}{S\arabic{equation}}
\renewcommand{\thefigure}{S\arabic{figure}}
\renewcommand{\bibnumfmt}[1]{[S#1]}
\renewcommand{\citenumfont}[1]{S#1}

%%%%%%%%%%%%%%%%%%%%%%%%%%%%%%%%%%%%%%%%%%%%%%%%%%%%%%%%%%%%%%%%%%%%%%%%%%%%%%%%%%%%%%%%%%%%%%%%%%%%%
% Experimental details
%%%%%%%%%%%%%%%%%%%%%%%%%%%%%%%%%%%%%%%%%%%%%%%%%%%%%%%%%%%%%%%%%%%%%%%%%%%%%%%%%%%%%%%%%%%%%%%%%%%%%
\section*{S1. Sample preparation and experimental procedures}

Polycrystalline spherical samples of Ni$_{1-x}$V$_x$ with V concentrations $x=0$ to 15\% were prepared by arc melting from
high purity elements (Ni 99.995\%, Ni$^{58}$ 99.9\%, V 99.8\%), annealed at 900 - 1050$^\circ$C for 3 days, cooled rapidly ($>200^{\circ}$C/min) and characterized as described in Refs.\
\cite{UbaidKassisVojtaSchroeder10sup,Schroederetal14sup}.

Magnetization and ac-susceptibility were measured in a Quantum Design SQUID magnetometer and in an
Oxford $^3$He/$^4$He dilution or $^4$He cryostat equipped with a pickup coil. A small orbital contribution has been subtracted
from the magnetization as explained in \cite{UbaidKassisVojtaSchroeder10sup}, and all data (except the ac susceptibility)
are demagnetized displaying the internal field $H$.

Muon spin rotation ($\mu$SR) data were collected at the DOLLY instrument at Swiss Muon Source (S$\mu$S), Paul Scherrer Institut and at the MuSR instrument at the ISIS facility using 7-30 pellets of each composition wrapped in silver foil.
All samples
%shown here
were measured at DOLLY in a similar Ag-mount,
the
%paramagnetic
compound
$x=12.3\%$ was also investigated in different cryostats at MuSR.
%for longer times and lower temperatures.
The asymmetry is shown for the DOLLY data, the other data (with different background) are corrected to match.

To probe the structure and chemical distribution, neutron diffraction data of the same samples were collected \cite{Gebretsadiketal_unpublishedsup} at the NPDF instrument \cite{NPDFsup} at the Los Alamos Neutron Science Center and at the NOMAD instrument \cite{NOMADsup} at the Spallation Neutron Source.
A detailed pair distribution function analysis (using PDFgui \cite{PDFFITsup,*PDFguisup}) does not reveal any deviations from a FCC-lattice with random occupancy of V atoms, the fit quality for $x=15\%$ is as high as in pure Ni.
PDF data is known to be sensitive to presence and length-scale of chemical short range order \cite{Cu3Ausup,LMNOsup}.
While simulated neutron PDF patterns of several V aggregate models confirm a weak sensitivity to the presence of V,
V-cluster model fits \cite{Gebretsadiketal_unpublishedsup} to experimental data were found inferior to random occupancy model fits.  The lattice constant and the average atomic displacements increase with $x$ as expected due to the larger V-radius ($r_V/r_{Ni}\approx1.05$)\cite{Gebretsadiketal_unpublishedsup}.

%%%%%%%%%%%%%%%%%%%%%%%%%%%%%%%%%%%%%%%%%%%%%%%%%%%%%%%%%%%%%%%%%%%%%%%%%%%%%%%%%%%%%%%%%%%%%%%%%%%%%
% muSR
%%%%%%%%%%%%%%%%%%%%%%%%%%%%%%%%%%%%%%%%%%%%%%%%%%%%%%%%%%%%%%%%%%%%%%%%%%%%%%%%%%%%%%%%%%%%%%%%%%%%%
\section*{S2. Raw data and details of the muon spin rotation analysis}

\begin{figure}
\includegraphics[width=16pc]{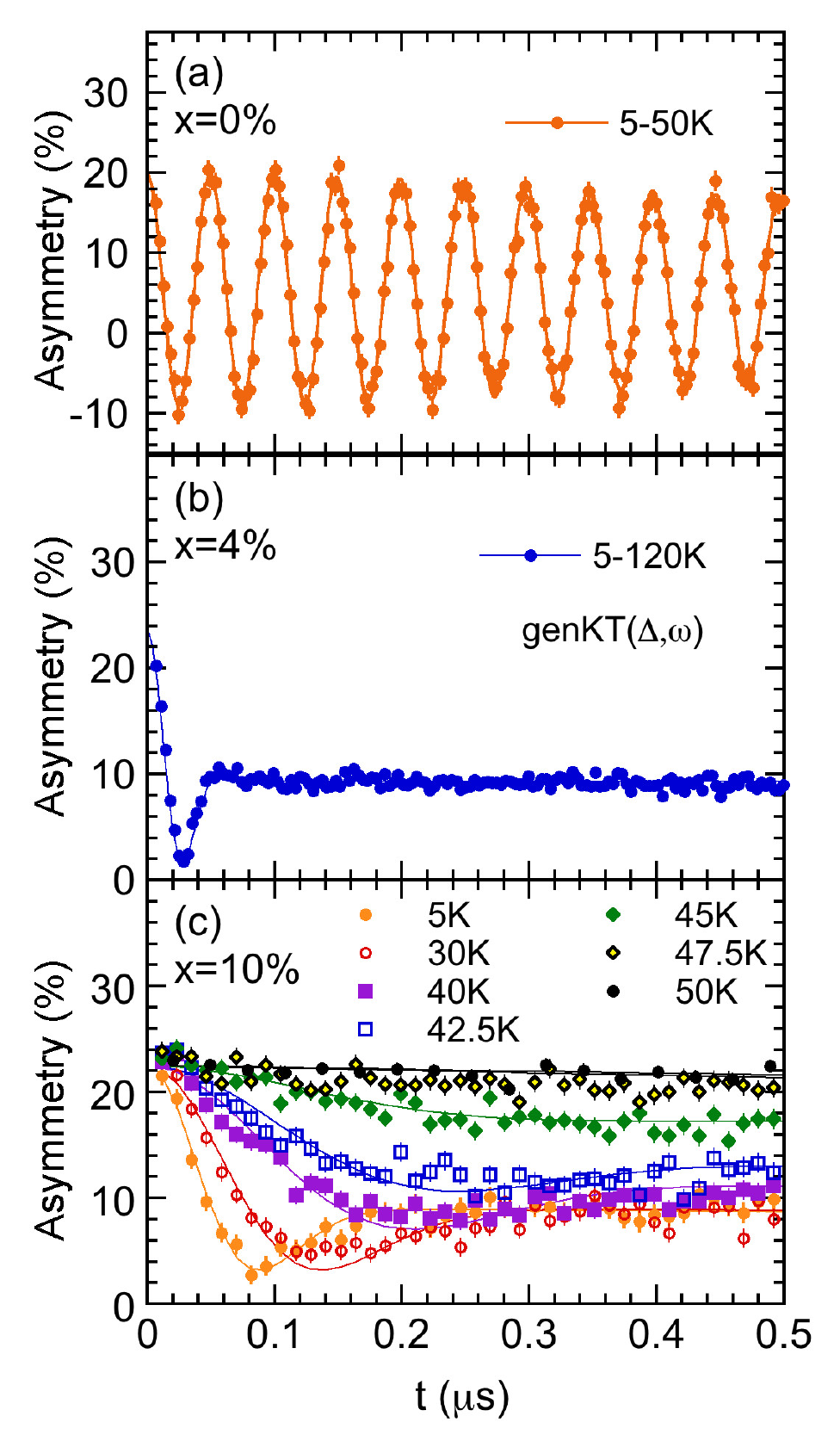}% Fig1%%%%%%%%%%%
\caption{Zero-field $\mu$SR asymmetry $A$ vs.\ time $t$  of Ni$_{1-x}$V$_x$.
(a) for $x=0$, (b) for $x=4\%$ and (c) for $x=10\%$ at low temperatures $T$.
Lines represent fits to Eq.\ (\ref{eq:Asum}) with  $P_{FM}=P_{genKT}$ for $x=0\%,4\%$ and with $P_{FM}=P_{KT}$ for $x=10\%$. Parameters shown in Fig. \ref{fig:x} and in Fig. \ref{fig:param}
 }
\label{fig:A010}
\end{figure}

\begin{figure}
\includegraphics[width=16pc]{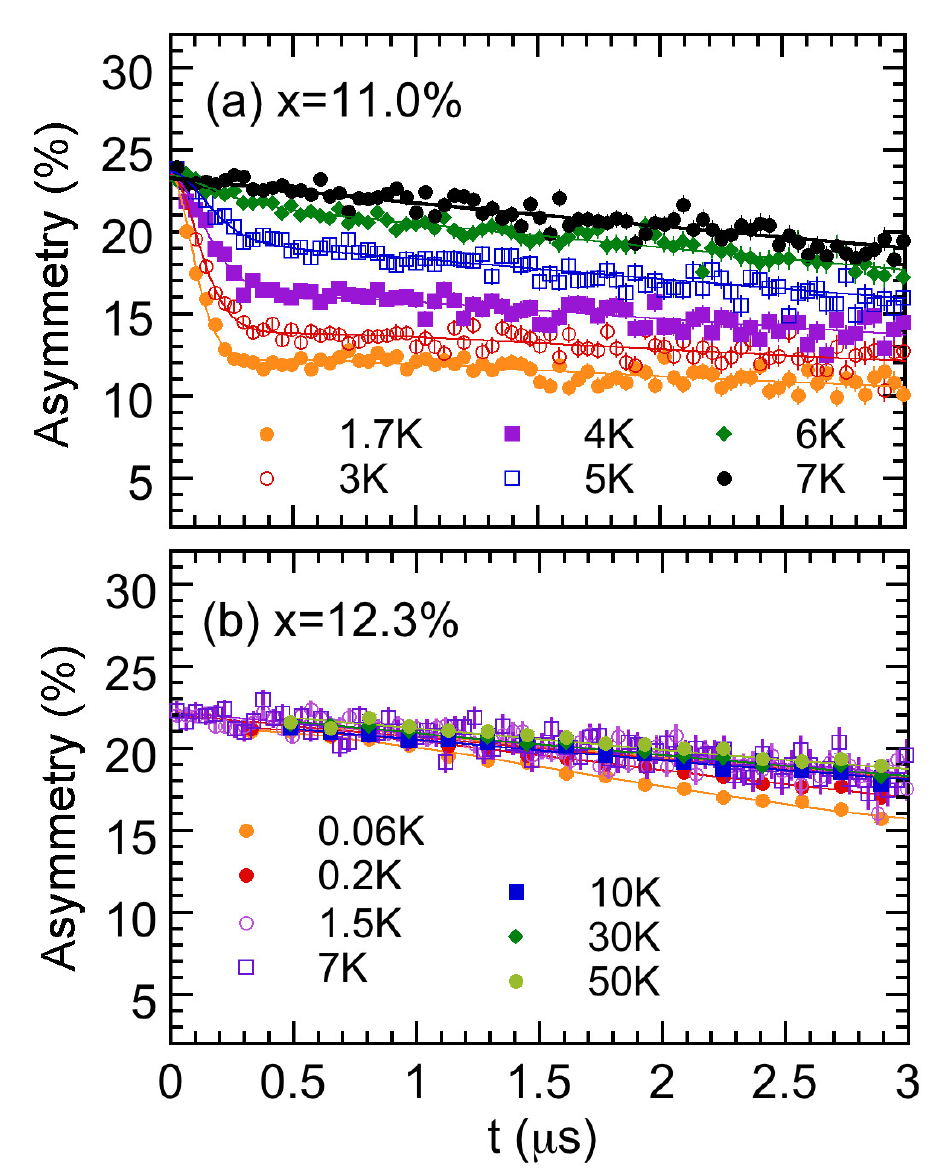}% Fig2 %%%%%%%%%
\caption{ Zero-field $\mu$SR asymmetry $A$ vs.\ time $t$ for different concentrations $x$ and temperatures $T$. (a) $A(t)$ for $x=11.0\%$.
Lines represent fits to Eq.\ (\ref{eq:Asum}) with $P_{FM}=P_{GBG}$. Parameters shown in Fig. \ref{fig:param}.
(b) $A(t)$ for $x=12.3\%$. Lines follow $P_{PM}$.
 }
\label{fig:A1112}
\end{figure}

The muon asymmetries $A(t)$ in zero magnetic field for samples having compositions $x=0$, 4\%, 10\%, 11\%, and 12.3\%
are shown in Figs.\ \ref{fig:A010} and \ref{fig:A1112}.
%figs here before
We model these data over the entire composition and temperature range by a superposition
of a ferromagnetic component with amplitude $A_{FM}$, a paramagnetic component with amplitude $A_{PM}$ and
a small constant background $A_{BG}$,
\begin{equation}
A(t)=A_{FM}(T) P_{FM}+A_{PM}(T) P_{PM}+A_{BG}~.
\label{eq:Asum}
\end{equation}
$A_{BG}$ is mainly due to the silver mount and was about $2\%$ for all Ni$_{1-x}$V$_x$ samples (measured in DOLLY instrument).
The paramagnetic component is well described by simple exponential decay,
\begin{equation}
P_{PM}=P_{exp}=exp(-\lambda t).
\end{equation}
It is dominated by the quasistatic nuclear V-spins and also contains the effects of fluctuating Ni magnetic clusters.
For the nuclear spins, the exponential decay can be understood as the short term limit of a Lorentzian Kubo-Toyabe
form appropriate for the diluted V atoms. The fluctuations of Ni magnetic clusters are expected to be fast
in this d-metal system with high $T_c(x=0)=630\,\textrm{K}$. The average decay rate $\lambda$ is thus very small and well in the motional narrowing regime.

The ferromagnetic component is modeled by different functional forms depending on the composition $x$
(see e.g. overview in Ref.\ \cite{Schroederetal14sup}).
For $x=0$ and 4\%, we use a generalized Kubo-Toyabe \cite{Hayanoetal79sup} form $P_{genKT}$ \cite{Kornilovsup,YaouancReotier11sup},
\begin{equation}
P_{genKT}
= \frac{1}{3} +\frac{2}{3} \exp(-\frac{\Delta^2 t^2}{2}) [\cos(\omega t)-\frac{\Delta^2 t}{\omega}\sin(\omega t)]~.
\end{equation}
It corresponds to a Gaussian distribution of the local magnetic fields with a nonzero mean $H_0=\omega/\gamma_{\mu} \neq0$
and width $\Delta H=\Delta/\gamma_{\mu}$ with $\gamma_{\mu}=2\pi \times135.5$\,MHz/T.
If $\Delta=0$, this function reduces to a simple undamped oscillation indicating a homogeneous field (at the same stopping site) in a uniform ferromagnet, as observed for $x=0$.
For $x=7-10$\%, we use a static Kubo-Toyabe (KT) \cite{Hayanoetal79sup} function
\begin{equation}
P_{KT}(t; \Delta,\nu =0) =\frac{1}{3} +\frac{2}{3} (1-\Delta^2 t^2) \exp(-\frac{1}{2}\Delta^2 t^2) \\
\end{equation}
at low temperatures.
It indicates a Gaussian distribution of local fields with negligible mean and width $\Delta H=\Delta/\gamma_{\mu}$.
At temperatures closer to $T_c$, a dynamic KT form with a finite fluctuation rate $\nu$ provides a better fit (using the WIMDA program \cite{Pratt2000sup}).

The KT form works well for $x$ up to 10\% but fails for $x\geq11.0\%$. (Note that longitudinal-field data do not support
a dynamic KT form for these samples \cite{Schroederetal14sup}.) A
better description for $x=11\%$ is achieved by using a broader static field distribution.
A Gaussian broadened Gaussian (GBG) is the superposition of multiple Gaussians whose widths
themselves are Gaussian distributed with mean $\Delta_0$ and width $w$. It takes the form
\cite{NoakesKalvius97sup}
\begin{equation}
P_{GBG}= \frac{1}{3}+\frac{2}{3} (\frac{1+R^2}{N})^{3/2}(1-\frac{\Delta^2 t^2}{N} \exp(-\frac{\Delta^2 t^2}{2N})
\end{equation}
with $N=1+R^2+R^2\Delta^2t^2$, where $R=w/\Delta_0$ is a relative distribution width and $\Delta^2=\Delta_{GBG}^2=\Delta_0^2+w^2$ is the square of the recorded effective width.
To account for dynamics, the longitudinal term was multiplied by an exponential giving
\begin{equation}
P_{GBG}(t;\nu)= P_{GBG}(t;\nu=0)+\frac{1}{3} (\exp(-\frac{2}{3}\nu t)-1)
\end{equation}
where $\nu$ is the fluctuation rate of the field.

\begin{figure}
\includegraphics[width=16pc]{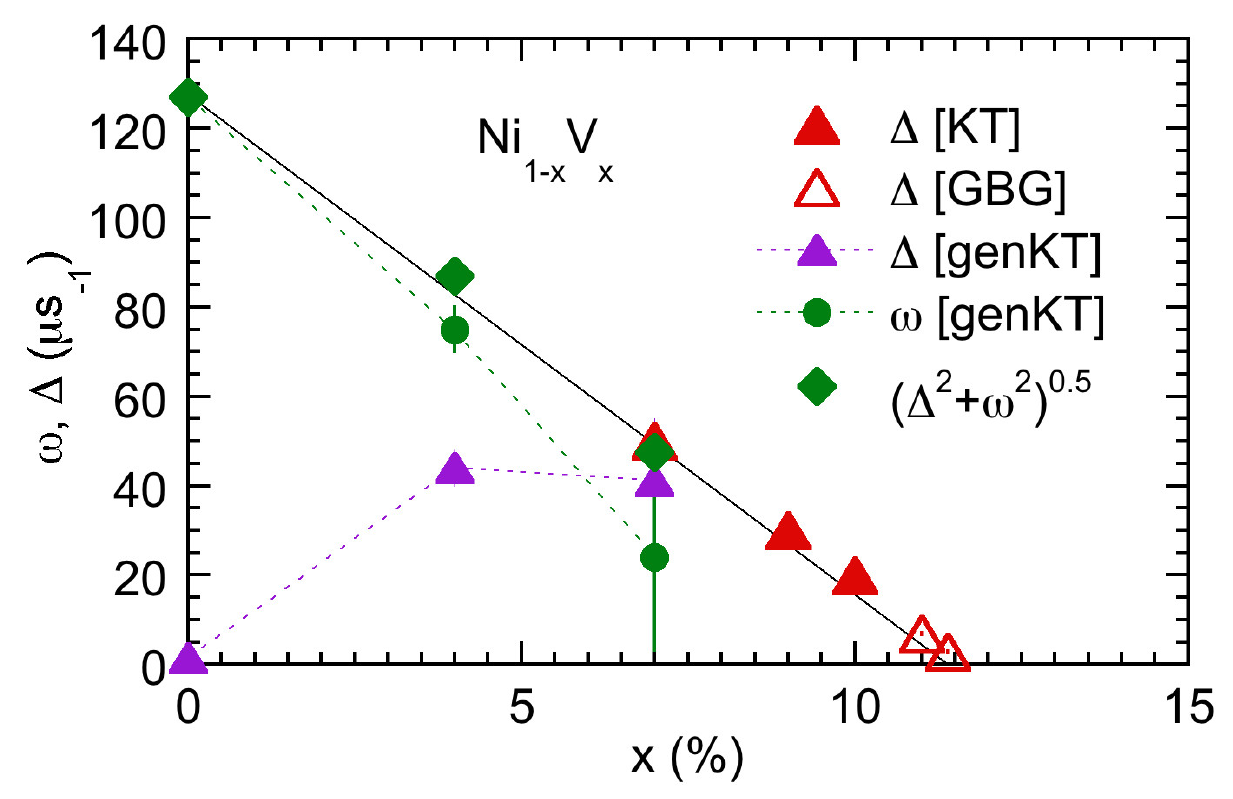}% Fig3 for intro overview
\caption{Energy scales characterizing the local field distribution for various $P_{FM}$ models vs.\ V-concentration $x$}
\label{fig:x}
\end{figure}

The evolution of the mean and width of the local-field distribution with $x$ at low temperatures is shown in Fig. \ref{fig:x}. The average field $\omega/\gamma_{\mu}$ decreases rapidly with $x$ as the probability of large domains is reduced beyond $x_c/2$, while the width $\Delta/\gamma_{\mu}$ shows a maximum at about $x_c/2$. An effective field $\sqrt{\Delta^2+\omega^2}/\gamma_{\mu}$ is linearly suppressed with $x$ and can be scaled directly to the mean magnetic moment $m_s(x)$ \cite{Schroederetal14sup}.

\begin{figure}
\includegraphics[width=20pc]{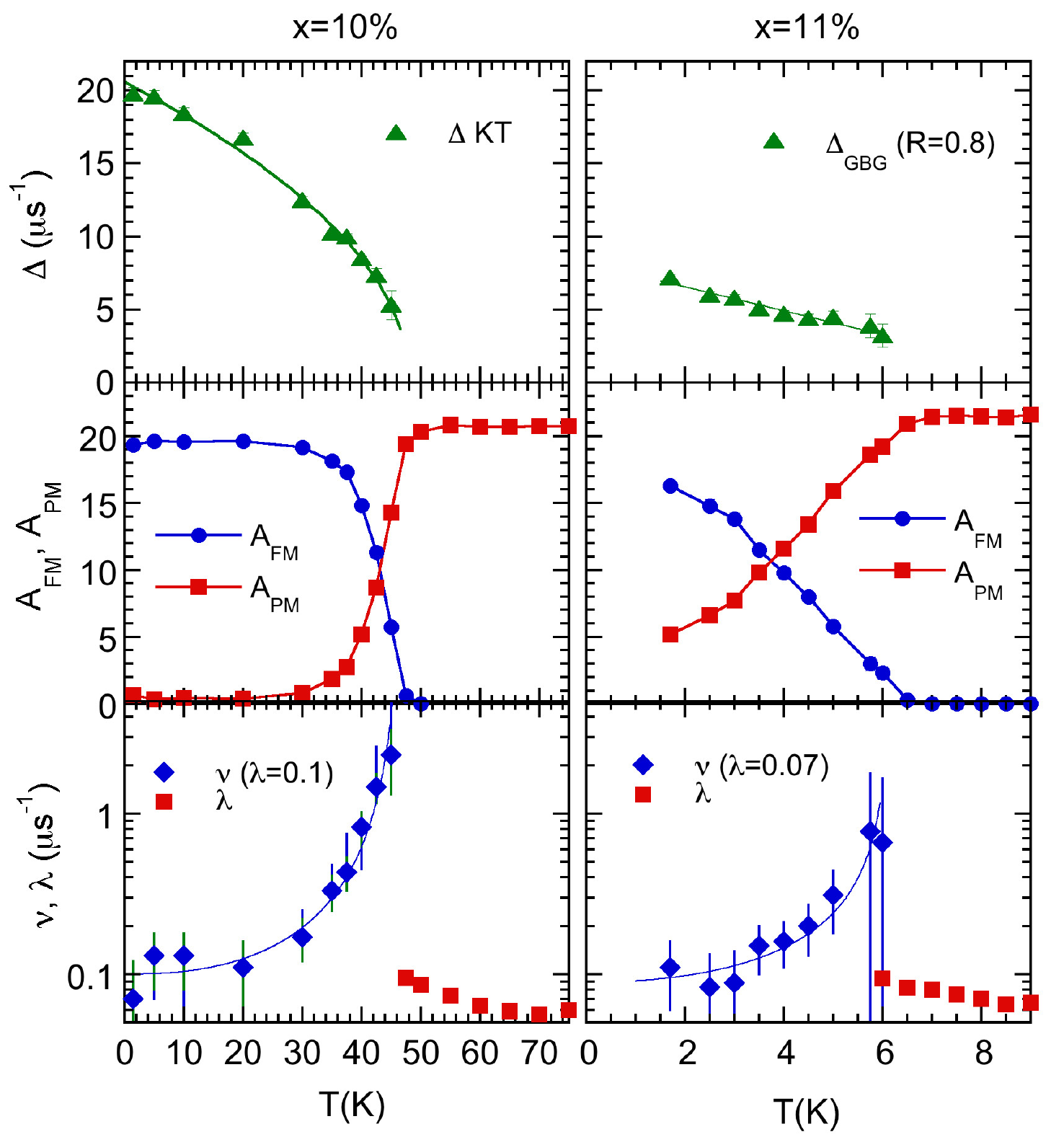}% Fig4  %%%%%%%%%
\caption{
Temperature dependence of the parameters in $\mu$SR analysis for $x=10\%$ (left) and $x=11\%$ (right). Top row: KT distribution width $\Delta$ for $x=10\%$ and effective GBG width $\Delta=\Delta_{GBG}$ for $x=11\%$.
%in frequency units $\Delta=\gamma_{\mu} \langle B^2_{loc}\rangle^{1/2}$ (with $\gamma_{\mu}=2\pi \times135.5$\,MHz/T).
Middle row: ferromagnetic and paramagnetic amplitudes $A_{FM}$ and $A_{PM}$.
Bottom row: fluctuation rate $\nu$ in the FM phase and depolarization rate $\lambda$ of the PM phase.}
\label{fig:param}
\end{figure}

Figure \ref{fig:param} shows additional $\mu$SR details for the samples analyzed in the main text,  $x=10\%$ and $11\%$.
For $x=10\%$, a single Gaussian ($R=0$) is sufficient to represent the local field distribution. Its width $\Delta$ vanishes at $T_c$, defined by the vanishing of  $A_{FM}$ (which matches $T_c$ determined from thermodynamic measurements \cite{Wangetal15sup}). The FM contribution reaches nearly $100\%$ at low $T$. In contrast, for $x=11\%$, a broad field distribution ($R=0.8$ used for all $T$) is required to describe the increased inhomogeneities close to $x_c$; and the effective distribution width $\Delta$ does not vanish at $T_c$. Even at the lowest temperatures, the FM contribution does not reach 100\%, instead about 20\% of $A(t)$ remains paramagnetic.
In both compounds the FM contribution is essentially static at low $T$ (fluctuation rate $\nu<0.2\mu s^{-1}$); it becomes more dynamic towards $T_c$. The decay rate $\lambda$ of the PM component increases towards the phase transition indicating that it is sensitive towards the
magnetic fluctuations of the Ni spins and clusters.

We note in passing that the behavior of the  $x=11\%$ sample close to $T_c$ can also be described by a stretched exponential form $\exp(-(\lambda t)^{\beta})$ with $\beta \approx 0.5$. Such behavior is often found for spin glasses \cite{uemuraSGsup} or in non-Fermi liquid compounds \cite {McLaughlin04sup} and implies multiple time scales and/or spatial correlations. At lower temperatures $(T<T_c)$, a single stretched exponential form cannot account for the short and long time depolarization. The very different depolarization rates are better represented by a two-component model (\ref{eq:Asum}).

Our analysis of the $\mu$SR data thus demonstrates that the $x=11\%$ sample features strongly inhomogeneous static ferromagnetic order coexisting with fluctuating magnetic clusters.

%\bibliographystyle{apsrev4-1}
%\bibliography{../00Bibtex/rareregions}
%\bibliography{citations_tv_5mod}

\begin{thebibliography}{52}%
\makeatletter
\providecommand \@ifxundefined [1]{%
 \@ifx{#1\undefined}
}%
\providecommand \@ifnum [1]{%
 \ifnum #1\expandafter \@firstoftwo
 \else \expandafter \@secondoftwo
 \fi
}%
\providecommand \@ifx [1]{%
 \ifx #1\expandafter \@firstoftwo
 \else \expandafter \@secondoftwo
 \fi
}%
\providecommand \natexlab [1]{#1}%
\providecommand \enquote  [1]{``#1''}%
\providecommand \bibnamefont  [1]{#1}%
\providecommand \bibfnamefont [1]{#1}%
\providecommand \citenamefont [1]{#1}%
\providecommand \href@noop [0]{\@secondoftwo}%
\providecommand \href [0]{\begingroup \@sanitize@url \@href}%
\providecommand \@href[1]{\@@startlink{#1}\@@href}%
\providecommand \@@href[1]{\endgroup#1\@@endlink}%
\providecommand \@sanitize@url [0]{\catcode `\\12\catcode `\$12\catcode
  `\&12\catcode `\#12\catcode `\^12\catcode `\_12\catcode `\%12\relax}%
\providecommand \@@startlink[1]{}%
\providecommand \@@endlink[0]{}%
\providecommand \url  [0]{\begingroup\@sanitize@url \@url }%
\providecommand \@url [1]{\endgroup\@href {#1}{\urlprefix }}%
\providecommand \urlprefix  [0]{URL }%
\providecommand \Eprint [0]{\href }%
\providecommand \doibase [0]{http://dx.doi.org/}%
\providecommand \selectlanguage [0]{\@gobble}%
\providecommand \bibinfo  [0]{\@secondoftwo}%
\providecommand \bibfield  [0]{\@secondoftwo}%
\providecommand \translation [1]{[#1]}%
\providecommand \BibitemOpen [0]{}%
\providecommand \bibitemStop [0]{}%
\providecommand \bibitemNoStop [0]{.\EOS\space}%
\providecommand \EOS [0]{\spacefactor3000\relax}%
\providecommand \BibitemShut  [1]{\csname bibitem#1\endcsname}%
\let\auto@bib@innerbib\@empty
%</preamble>
\bibitem [{\citenamefont {Sachdev}(1999)}]{Sachdev99}%
  \BibitemOpen
  \bibfield  {author} {\bibinfo {author} {\bibfnamefont {S.}~\bibnamefont
  {Sachdev}},\ }\href@noop {} {\bibfield  {journal} {\bibinfo  {journal}
  {Physics World}\ }\textbf {\bibinfo {volume} {12}},\ \bibinfo {pages} {33}
  (\bibinfo {year} {1999})}\BibitemShut {NoStop}%
\bibitem [{\citenamefont {von L{\"o}hneysen}\ \emph {et~al.}(2007)\citenamefont
  {von L{\"o}hneysen}, \citenamefont {Rosch}, \citenamefont {Vojta},\ and\
  \citenamefont {W{\"o}lfle}}]{LRVW07}%
  \BibitemOpen
  \bibfield  {author} {\bibinfo {author} {\bibfnamefont {H.}~\bibnamefont {von
  L{\"o}hneysen}}, \bibinfo {author} {\bibfnamefont {A.}~\bibnamefont {Rosch}},
  \bibinfo {author} {\bibfnamefont {M.}~\bibnamefont {Vojta}}, \ and\ \bibinfo
  {author} {\bibfnamefont {P.}~\bibnamefont {W{\"o}lfle}},\ }\href@noop {}
  {\bibfield  {journal} {\bibinfo  {journal} {Rev. Mod. Phys.}\ }\textbf
  {\bibinfo {volume} {79}},\ \bibinfo {pages} {1015} (\bibinfo {year}
  {2007})}\BibitemShut {NoStop}%
\bibitem [{\citenamefont {Fisher}(1992)}]{Fisher92}%
  \BibitemOpen
  \bibfield  {author} {\bibinfo {author} {\bibfnamefont {D.~S.}\ \bibnamefont
  {Fisher}},\ }\href@noop {} {\bibfield  {journal} {\bibinfo  {journal} {Phys.
  Rev. Lett.}\ }\textbf {\bibinfo {volume} {69}},\ \bibinfo {pages} {534}
  (\bibinfo {year} {1992})}\BibitemShut {NoStop}%
\bibitem [{\citenamefont {Fisher}(1995)}]{Fisher95}%
  \BibitemOpen
  \bibfield  {author} {\bibinfo {author} {\bibfnamefont {D.~S.}\ \bibnamefont
  {Fisher}},\ }\href {\doibase 10.1103/PhysRevB.51.6411} {\bibfield  {journal}
  {\bibinfo  {journal} {Phys. Rev. B}\ }\textbf {\bibinfo {volume} {51}},\
  \bibinfo {pages} {6411} (\bibinfo {year} {1995})}\BibitemShut {NoStop}%
\bibitem [{\citenamefont {Thill}\ and\ \citenamefont
  {Huse}(1995)}]{ThillHuse95}%
  \BibitemOpen
  \bibfield  {author} {\bibinfo {author} {\bibfnamefont {M.}~\bibnamefont
  {Thill}}\ and\ \bibinfo {author} {\bibfnamefont {D.~A.}\ \bibnamefont
  {Huse}},\ }\href {\doibase 10.1016/0378-4371(94)00247-Q} {\bibfield
  {journal} {\bibinfo  {journal} {Physica A}\ }\textbf {\bibinfo {volume}
  {214}},\ \bibinfo {pages} {321} (\bibinfo {year} {1995})}\BibitemShut
  {NoStop}%
\bibitem [{\citenamefont {Castro~Neto}\ and\ \citenamefont
  {Jones}(2000)}]{CastroNetoJones00}%
  \BibitemOpen
  \bibfield  {author} {\bibinfo {author} {\bibfnamefont {A.~H.}\ \bibnamefont
  {Castro~Neto}}\ and\ \bibinfo {author} {\bibfnamefont {B.~A.}\ \bibnamefont
  {Jones}},\ }\href@noop {} {\bibfield  {journal} {\bibinfo  {journal} {Phys.
  Rev. B}\ }\textbf {\bibinfo {volume} {62}},\ \bibinfo {pages} {14975}
  (\bibinfo {year} {2000})}\BibitemShut {NoStop}%
\bibitem [{\citenamefont {Vojta}\ and\ \citenamefont
  {Schmalian}(2005)}]{VojtaSchmalian05}%
  \BibitemOpen
  \bibfield  {author} {\bibinfo {author} {\bibfnamefont {T.}~\bibnamefont
  {Vojta}}\ and\ \bibinfo {author} {\bibfnamefont {J.}~\bibnamefont
  {Schmalian}},\ }\href {\doibase 10.1103/PhysRevB.72.045438} {\bibfield
  {journal} {\bibinfo  {journal} {Phys. Rev. B}\ }\textbf {\bibinfo {volume}
  {72}},\ \bibinfo {pages} {045438} (\bibinfo {year} {2005})}\BibitemShut
  {NoStop}%
\bibitem [{\citenamefont {Hoyos}\ \emph {et~al.}(2007)\citenamefont {Hoyos},
  \citenamefont {Kotabage},\ and\ \citenamefont
  {Vojta}}]{HoyosKotabageVojta07}%
  \BibitemOpen
  \bibfield  {author} {\bibinfo {author} {\bibfnamefont {J.~A.}\ \bibnamefont
  {Hoyos}}, \bibinfo {author} {\bibfnamefont {C.}~\bibnamefont {Kotabage}}, \
  and\ \bibinfo {author} {\bibfnamefont {T.}~\bibnamefont {Vojta}},\ }\href
  {\doibase 10.1103/PhysRevLett.99.230601} {\bibfield  {journal} {\bibinfo
  {journal} {Phys. Rev. Lett.}\ }\textbf {\bibinfo {volume} {99}},\ \bibinfo
  {pages} {230601} (\bibinfo {year} {2007})}\BibitemShut {NoStop}%
\bibitem [{\citenamefont {Vojta}\ \emph {et~al.}(2009)\citenamefont {Vojta},
  \citenamefont {Kotabage},\ and\ \citenamefont
  {Hoyos}}]{VojtaKotabageHoyos09}%
  \BibitemOpen
  \bibfield  {author} {\bibinfo {author} {\bibfnamefont {T.}~\bibnamefont
  {Vojta}}, \bibinfo {author} {\bibfnamefont {C.}~\bibnamefont {Kotabage}}, \
  and\ \bibinfo {author} {\bibfnamefont {J.~A.}\ \bibnamefont {Hoyos}},\ }\href
  {\doibase 10.1103/PhysRevB.79.024401} {\bibfield  {journal} {\bibinfo
  {journal} {Phys. Rev. B}\ }\textbf {\bibinfo {volume} {79}},\ \bibinfo
  {pages} {024401} (\bibinfo {year} {2009})}\BibitemShut {NoStop}%
\bibitem [{\citenamefont {Del~Maestro}\ \emph {et~al.}(2008)\citenamefont
  {Del~Maestro}, \citenamefont {Rosenow}, \citenamefont {M{\"u}ller},\ and\
  \citenamefont {Sachdev}}]{DRMS08}%
  \BibitemOpen
  \bibfield  {author} {\bibinfo {author} {\bibfnamefont {A.}~\bibnamefont
  {Del~Maestro}}, \bibinfo {author} {\bibfnamefont {B.}~\bibnamefont
  {Rosenow}}, \bibinfo {author} {\bibfnamefont {M.}~\bibnamefont {M{\"u}ller}},
  \ and\ \bibinfo {author} {\bibfnamefont {S.}~\bibnamefont {Sachdev}},\
  }\href@noop {} {\bibfield  {journal} {\bibinfo  {journal} {Phys. Rev. Lett.}\
  }\textbf {\bibinfo {volume} {101}},\ \bibinfo {pages} {035701} (\bibinfo
  {year} {2008})}\BibitemShut {NoStop}%
\bibitem [{\citenamefont {Vojta}(2006)}]{Vojta06}%
  \BibitemOpen
  \bibfield  {author} {\bibinfo {author} {\bibfnamefont {T.}~\bibnamefont
  {Vojta}},\ }\href {\doibase 10.1088/0305-4470/39/22/R01} {\bibfield
  {journal} {\bibinfo  {journal} {J. Phys. A}\ }\textbf {\bibinfo {volume}
  {39}},\ \bibinfo {pages} {R143} (\bibinfo {year} {2006})}\BibitemShut
  {NoStop}%
\bibitem [{\citenamefont {Vojta}(2010)}]{Vojta10}%
  \BibitemOpen
  \bibfield  {author} {\bibinfo {author} {\bibfnamefont {T.}~\bibnamefont
  {Vojta}},\ }\href {\doibase 10.1007/s10909-010-0205-4} {\bibfield  {journal}
  {\bibinfo  {journal} {J. Low Temp. Phys.}\ }\textbf {\bibinfo {volume}
  {161}},\ \bibinfo {pages} {299} (\bibinfo {year} {2010})}\BibitemShut
  {NoStop}%
\bibitem [{\citenamefont {Sereni}\ \emph {et~al.}(2007)\citenamefont {Sereni},
  \citenamefont {Westerkamp}, \citenamefont {K{\"u}chler}, \citenamefont
  {Caroca-Canales}, \citenamefont {Gegenwart},\ and\ \citenamefont
  {Geibel}}]{SWKCGG07}%
  \BibitemOpen
  \bibfield  {author} {\bibinfo {author} {\bibfnamefont {J.~G.}\ \bibnamefont
  {Sereni}}, \bibinfo {author} {\bibfnamefont {T.}~\bibnamefont {Westerkamp}},
  \bibinfo {author} {\bibfnamefont {R.}~\bibnamefont {K{\"u}chler}}, \bibinfo
  {author} {\bibfnamefont {N.}~\bibnamefont {Caroca-Canales}}, \bibinfo
  {author} {\bibfnamefont {P.}~\bibnamefont {Gegenwart}}, \ and\ \bibinfo
  {author} {\bibfnamefont {C.}~\bibnamefont {Geibel}},\ }\href@noop {}
  {\bibfield  {journal} {\bibinfo  {journal} {Phys. Rev. B}\ }\textbf {\bibinfo
  {volume} {75}},\ \bibinfo {pages} {024432} (\bibinfo {year}
  {2007})}\BibitemShut {NoStop}%
\bibitem [{\citenamefont {Westerkamp}\ \emph {et~al.}(2009)\citenamefont
  {Westerkamp}, \citenamefont {Deppe}, \citenamefont {K{\"u}chler},
  \citenamefont {Brando}, \citenamefont {Geibel}, \citenamefont {Gegenwart},
  \citenamefont {Pikul},\ and\ \citenamefont {Steglich}}]{Westerkampetal09}%
  \BibitemOpen
  \bibfield  {author} {\bibinfo {author} {\bibfnamefont {T.}~\bibnamefont
  {Westerkamp}}, \bibinfo {author} {\bibfnamefont {M.}~\bibnamefont {Deppe}},
  \bibinfo {author} {\bibfnamefont {R.}~\bibnamefont {K{\"u}chler}}, \bibinfo
  {author} {\bibfnamefont {M.}~\bibnamefont {Brando}}, \bibinfo {author}
  {\bibfnamefont {C.}~\bibnamefont {Geibel}}, \bibinfo {author} {\bibfnamefont
  {P.}~\bibnamefont {Gegenwart}}, \bibinfo {author} {\bibfnamefont {A.~P.}\
  \bibnamefont {Pikul}}, \ and\ \bibinfo {author} {\bibfnamefont
  {F.}~\bibnamefont {Steglich}},\ }\href@noop {} {\bibfield  {journal}
  {\bibinfo  {journal} {Phys. Rev. Lett.}\ }\textbf {\bibinfo {volume} {102}},\
  \bibinfo {pages} {206404} (\bibinfo {year} {2009})}\BibitemShut {NoStop}%
\bibitem [{\citenamefont {Ubaid-Kassis}\ \emph {et~al.}(2010)\citenamefont
  {Ubaid-Kassis}, \citenamefont {Vojta},\ and\ \citenamefont
  {Schroeder}}]{UbaidKassisVojtaSchroeder10}%
  \BibitemOpen
  \bibfield  {author} {\bibinfo {author} {\bibfnamefont {S.}~\bibnamefont
  {Ubaid-Kassis}}, \bibinfo {author} {\bibfnamefont {T.}~\bibnamefont {Vojta}},
  \ and\ \bibinfo {author} {\bibfnamefont {A.}~\bibnamefont {Schroeder}},\
  }\href@noop {} {\bibfield  {journal} {\bibinfo  {journal} {Phys. Rev. Lett.}\
  }\textbf {\bibinfo {volume} {104}},\ \bibinfo {pages} {066402} (\bibinfo
  {year} {2010})}\BibitemShut {NoStop}%
\bibitem [{\citenamefont {Schroeder}\ \emph {et~al.}(2011)\citenamefont
  {Schroeder}, \citenamefont {Ubaid-Kassis},\ and\ \citenamefont
  {Vojta}}]{SchroederUbaidKassisVojta11}%
  \BibitemOpen
  \bibfield  {author} {\bibinfo {author} {\bibfnamefont {A.}~\bibnamefont
  {Schroeder}}, \bibinfo {author} {\bibfnamefont {S.}~\bibnamefont
  {Ubaid-Kassis}}, \ and\ \bibinfo {author} {\bibfnamefont {T.}~\bibnamefont
  {Vojta}},\ }\href@noop {} {\bibfield  {journal} {\bibinfo  {journal} {J.
  Phys. Condens. Matter}\ }\textbf {\bibinfo {volume} {23}},\ \bibinfo {pages}
  {094205} (\bibinfo {year} {2011})}\BibitemShut {NoStop}%
\bibitem [{\citenamefont {Senthil}\ and\ \citenamefont
  {Sachdev}(1996)}]{SenthilSachdev96}%
  \BibitemOpen
  \bibfield  {author} {\bibinfo {author} {\bibfnamefont {T.}~\bibnamefont
  {Senthil}}\ and\ \bibinfo {author} {\bibfnamefont {S.}~\bibnamefont
  {Sachdev}},\ }\href {\doibase 10.1103/PhysRevLett.77.5292} {\bibfield
  {journal} {\bibinfo  {journal} {Phys. Rev. Lett.}\ }\textbf {\bibinfo
  {volume} {77}},\ \bibinfo {pages} {5292} (\bibinfo {year}
  {1996})}\BibitemShut {NoStop}%
\bibitem [{\citenamefont {Motrunich}\ \emph {et~al.}(2000)\citenamefont
  {Motrunich}, \citenamefont {Mau}, \citenamefont {Huse},\ and\ \citenamefont
  {Fisher}}]{MMHF00}%
  \BibitemOpen
  \bibfield  {author} {\bibinfo {author} {\bibfnamefont {O.}~\bibnamefont
  {Motrunich}}, \bibinfo {author} {\bibfnamefont {S.~C.}\ \bibnamefont {Mau}},
  \bibinfo {author} {\bibfnamefont {D.~A.}\ \bibnamefont {Huse}}, \ and\
  \bibinfo {author} {\bibfnamefont {D.~S.}\ \bibnamefont {Fisher}},\ }\href
  {\doibase 10.1103/PhysRevB.61.1160} {\bibfield  {journal} {\bibinfo
  {journal} {Phys. Rev. B}\ }\textbf {\bibinfo {volume} {61}},\ \bibinfo
  {pages} {1160} (\bibinfo {year} {2000})}\BibitemShut {NoStop}%
\bibitem [{Note1()}]{Note1}%
  \BibitemOpen
  \bibinfo {note} {Unusual scaling behavior in the ferromagnetic phase of
  URu$_{2-x}$Re$_x$Si$_2$ was initially suggested to stem from a Griffiths
  phase but later work showed that this is likely not the case \cite
  {Baueretal05,ButchMaple09}.}\BibitemShut {Stop}%
\bibitem [{\citenamefont {Brando}\ \emph {et~al.}(2016)\citenamefont {Brando},
  \citenamefont {Belitz}, \citenamefont {Grosche},\ and\ \citenamefont
  {Kirkpatrick}}]{BrandoBelitzGroscheKirkpatrick16}%
  \BibitemOpen
  \bibfield  {author} {\bibinfo {author} {\bibfnamefont {M.}~\bibnamefont
  {Brando}}, \bibinfo {author} {\bibfnamefont {D.}~\bibnamefont {Belitz}},
  \bibinfo {author} {\bibfnamefont {F.~M.}\ \bibnamefont {Grosche}}, \ and\
  \bibinfo {author} {\bibfnamefont {T.~R.}\ \bibnamefont {Kirkpatrick}},\
  }\href {\doibase 10.1103/RevModPhys.88.025006} {\bibfield  {journal}
  {\bibinfo  {journal} {Rev. Mod. Phys.}\ }\textbf {\bibinfo {volume} {88}},\
  \bibinfo {pages} {025006} (\bibinfo {year} {2016})}\BibitemShut {NoStop}%
\bibitem [{\citenamefont {Schroeder}\ \emph {et~al.}(2014)\citenamefont
  {Schroeder}, \citenamefont {Wang}, \citenamefont {Baker}, \citenamefont
  {Pratt}, \citenamefont {Blundell}, \citenamefont {Lancaster}, \citenamefont
  {Franke},\ and\ \citenamefont {M\"oller}}]{Schroederetal14}%
  \BibitemOpen
  \bibfield  {author} {\bibinfo {author} {\bibfnamefont {A.}~\bibnamefont
  {Schroeder}}, \bibinfo {author} {\bibfnamefont {R.}~\bibnamefont {Wang}},
  \bibinfo {author} {\bibfnamefont {P.~J.}\ \bibnamefont {Baker}}, \bibinfo
  {author} {\bibfnamefont {F.~L.}\ \bibnamefont {Pratt}}, \bibinfo {author}
  {\bibfnamefont {S.~J.}\ \bibnamefont {Blundell}}, \bibinfo {author}
  {\bibfnamefont {T.}~\bibnamefont {Lancaster}}, \bibinfo {author}
  {\bibfnamefont {I.}~\bibnamefont {Franke}}, \ and\ \bibinfo {author}
  {\bibfnamefont {J.~S.}\ \bibnamefont {M\"oller}},\ }\href
  {http://stacks.iop.org/1742-6596/551/i=1/a=012003} {\bibfield  {journal}
  {\bibinfo  {journal} {J. Phys. Conf. Series}\ }\textbf {\bibinfo {volume}
  {551}},\ \bibinfo {pages} {012003} (\bibinfo {year} {2014})}\BibitemShut
  {NoStop}%
\bibitem [{Sup()}]{Supplementary}%
  \BibitemOpen
  \href@noop {} {}\bibinfo {note} {See Supplemental Material in XXX with Refs.
  \cite{Gebretsadiketal_unpublished,NPDF,NOMAD,PDFFIT,*PDFgui,Cu3Au,LMNO,Kornilov,Pratt2000}
  for details on sample preparation, experiments and $\mu$SR
  analysis}\BibitemShut {NoStop}%
\bibitem [{\citenamefont {B\"{o}lling}(1968)}]{boelling68}%
  \BibitemOpen
  \bibfield  {author} {\bibinfo {author} {\bibfnamefont {F.}~\bibnamefont
  {B\"{o}lling}},\ }\href@noop {} {\bibfield  {journal} {\bibinfo  {journal}
  {Phys. Kondens. Mater.}\ }\textbf {\bibinfo {volume} {7}},\ \bibinfo {pages}
  {162} (\bibinfo {year} {1968})}\BibitemShut {NoStop}%
\bibitem [{\citenamefont {Gregory}\ and\ \citenamefont
  {Moody}(1975)}]{gregory75C}%
  \BibitemOpen
  \bibfield  {author} {\bibinfo {author} {\bibfnamefont {I.~P.}\ \bibnamefont
  {Gregory}}\ and\ \bibinfo {author} {\bibfnamefont {D.~E.}\ \bibnamefont
  {Moody}},\ }\href@noop {} {\bibfield  {journal} {\bibinfo  {journal} {J.
  Phys. F: Met. Phys.}\ }\textbf {\bibinfo {volume} {5}},\ \bibinfo {pages}
  {36} (\bibinfo {year} {1975})}\BibitemShut {NoStop}%
\bibitem [{\citenamefont {Nicklas}\ \emph {et~al.}(1999)\citenamefont
  {Nicklas}, \citenamefont {Brando}, \citenamefont {Knebel}, \citenamefont
  {Mayr}, \citenamefont {Trinkl},\ and\ \citenamefont {Loidl}}]{NBKMTL99}%
  \BibitemOpen
  \bibfield  {author} {\bibinfo {author} {\bibfnamefont {M.}~\bibnamefont
  {Nicklas}}, \bibinfo {author} {\bibfnamefont {M.}~\bibnamefont {Brando}},
  \bibinfo {author} {\bibfnamefont {G.}~\bibnamefont {Knebel}}, \bibinfo
  {author} {\bibfnamefont {F.}~\bibnamefont {Mayr}}, \bibinfo {author}
  {\bibfnamefont {W.}~\bibnamefont {Trinkl}}, \ and\ \bibinfo {author}
  {\bibfnamefont {A.}~\bibnamefont {Loidl}},\ }\href@noop {} {\bibfield
  {journal} {\bibinfo  {journal} {Phys. Rev. Lett.}\ }\textbf {\bibinfo
  {volume} {82}},\ \bibinfo {pages} {4268} (\bibinfo {year}
  {1999})}\BibitemShut {NoStop}%
\bibitem [{\citenamefont {Friedel}(1958)}]{Friedel58}%
  \BibitemOpen
  \bibfield  {author} {\bibinfo {author} {\bibfnamefont {J.}~\bibnamefont
  {Friedel}},\ }\href {\doibase 10.1007/BF02751483} {\bibfield  {journal}
  {\bibinfo  {journal} {Nuovo Cimento}\ }\textbf {\bibinfo {volume} {7}},\
  \bibinfo {pages} {287} (\bibinfo {year} {1958})}\BibitemShut {NoStop}%
\bibitem [{\citenamefont {Collins}\ and\ \citenamefont
  {Low}(1965)}]{CollinsLow65}%
  \BibitemOpen
  \bibfield  {author} {\bibinfo {author} {\bibfnamefont {M.~F.}\ \bibnamefont
  {Collins}}\ and\ \bibinfo {author} {\bibfnamefont {G.~G.}\ \bibnamefont
  {Low}},\ }\href {http://stacks.iop.org/0370-1328/86/i=3/a=313} {\bibfield
  {journal} {\bibinfo  {journal} {Proceedings of the Physical Society}\
  }\textbf {\bibinfo {volume} {86}},\ \bibinfo {pages} {535} (\bibinfo {year}
  {1965})}\BibitemShut {NoStop}%
\bibitem [{\citenamefont {Wang}\ \emph {et~al.}(2015)\citenamefont {Wang},
  \citenamefont {Ubaid-Kassis}, \citenamefont {Schroeder}, \citenamefont
  {Baker}, \citenamefont {Pratt}, \citenamefont {Blundell}, \citenamefont
  {Lancaster}, \citenamefont {Franke}, \citenamefont {M\"oller},\ and\
  \citenamefont {Vojta}}]{Wangetal15}%
  \BibitemOpen
  \bibfield  {author} {\bibinfo {author} {\bibfnamefont {R.}~\bibnamefont
  {Wang}}, \bibinfo {author} {\bibfnamefont {S.}~\bibnamefont {Ubaid-Kassis}},
  \bibinfo {author} {\bibfnamefont {A.}~\bibnamefont {Schroeder}}, \bibinfo
  {author} {\bibfnamefont {P.}~\bibnamefont {Baker}}, \bibinfo {author}
  {\bibfnamefont {F.}~\bibnamefont {Pratt}}, \bibinfo {author} {\bibfnamefont
  {S.}~\bibnamefont {Blundell}}, \bibinfo {author} {\bibfnamefont
  {T.}~\bibnamefont {Lancaster}}, \bibinfo {author} {\bibfnamefont
  {I.}~\bibnamefont {Franke}}, \bibinfo {author} {\bibfnamefont
  {J.}~\bibnamefont {M\"oller}}, \ and\ \bibinfo {author} {\bibfnamefont
  {T.}~\bibnamefont {Vojta}},\ }\href
  {http://stacks.iop.org/1742-6596/592/i=1/a=012089} {\bibfield  {journal}
  {\bibinfo  {journal} {J. Phys. Conf. Series}\ }\textbf {\bibinfo {volume}
  {592}},\ \bibinfo {pages} {012089} (\bibinfo {year} {2015})}\BibitemShut
  {NoStop}%
\bibitem [{Note2()}]{Note2}%
  \BibitemOpen
  \bibinfo {note} {This estimate of $\nu \psi $ is somewhat smaller than the
  value $\nu \psi =0.42$ obtained in Ref.\ \cite {UbaidKassisVojtaSchroeder10}
  from the $x$-dependence of the susceptibility exponent $\gamma $. The
  deviation is within the error bars.}\BibitemShut {Stop}%
\bibitem [{\citenamefont {Dobrosavljevic}\ and\ \citenamefont
  {Miranda}(2005)}]{DobrosavljevicMiranda05}%
  \BibitemOpen
  \bibfield  {author} {\bibinfo {author} {\bibfnamefont {V.}~\bibnamefont
  {Dobrosavljevic}}\ and\ \bibinfo {author} {\bibfnamefont {E.}~\bibnamefont
  {Miranda}},\ }\href@noop {} {\bibfield  {journal} {\bibinfo  {journal} {Phys.
  Rev. Lett.}\ }\textbf {\bibinfo {volume} {94}},\ \bibinfo {pages} {187203}
  (\bibinfo {year} {2005})}\BibitemShut {NoStop}%
\bibitem [{Note3()}]{Note3}%
  \BibitemOpen
  \bibinfo {note} {If the relevant rare regions are so large that order
  parameter conservation hampers the dynamics of the clusters, the functional
  form of the Griffiths singularities changes \cite {NozadzeVojta12}. Based on
  a typical cluster moment of about 12\protect \tmspace +\thinmuskip
  {.1667em}$\mu _B$ at $x_c$ \cite
  {UbaidKassisVojtaSchroeder10,SchroederUbaidKassisVojta11}, this likely does
  not play a role here.}\BibitemShut {Stop}%
\bibitem [{Note4()}]{Note4}%
  \BibitemOpen
  \bibinfo {note} {This measure somewhat overestimates the cluster contribution
  because $M_{hi}$ in $H=$ 50 kOe also contains the bulk response to the
  field.}\BibitemShut {Stop}%
\bibitem [{\citenamefont {Blundell}(1999)}]{Blundelleasy}%
  \BibitemOpen
  \bibfield  {author} {\bibinfo {author} {\bibfnamefont {S.~J.}\ \bibnamefont
  {Blundell}},\ }\href {\doibase 10.1080/001075199181521} {\bibfield  {journal}
  {\bibinfo  {journal} {Contemporary Physics}\ }\textbf {\bibinfo {volume}
  {40}},\ \bibinfo {pages} {175} (\bibinfo {year} {1999})}\BibitemShut
  {NoStop}%
\bibitem [{\citenamefont {Yaouanc}\ and\ \citenamefont {Dalmas~de
  Reotier}(2011)}]{YaouancReotier11}%
  \BibitemOpen
  \bibfield  {author} {\bibinfo {author} {\bibfnamefont {A.}~\bibnamefont
  {Yaouanc}}\ and\ \bibinfo {author} {\bibfnamefont {P.}~\bibnamefont
  {Dalmas~de Reotier}},\ }\href@noop {} {\emph {\bibinfo {title} {Muon Spin
  Rotation, Relaxation, and Resonance: Applications to Condensed Matter}}}\
  (\bibinfo  {publisher} {Oxford University Press},\ \bibinfo {address}
  {Oxford},\ \bibinfo {year} {2011})\BibitemShut {NoStop}%
\bibitem [{\citenamefont {Amato}(1997)}]{Amato97}%
  \BibitemOpen
  \bibfield  {author} {\bibinfo {author} {\bibfnamefont {A.}~\bibnamefont
  {Amato}},\ }\href {\doibase 10.1103/RevModPhys.69.1119} {\bibfield  {journal}
  {\bibinfo  {journal} {Rev. Mod. Phys.}\ }\textbf {\bibinfo {volume} {69}},\
  \bibinfo {pages} {1119} (\bibinfo {year} {1997})}\BibitemShut {NoStop}%
\bibitem [{\citenamefont {Uemura}\ \emph {et~al.}(1985)\citenamefont {Uemura},
  \citenamefont {Yamazaki}, \citenamefont {Harshman}, \citenamefont {Senba},\
  and\ \citenamefont {Ansaldo}}]{uemuraSG}%
  \BibitemOpen
  \bibfield  {author} {\bibinfo {author} {\bibfnamefont {Y.~J.}\ \bibnamefont
  {Uemura}}, \bibinfo {author} {\bibfnamefont {T.}~\bibnamefont {Yamazaki}},
  \bibinfo {author} {\bibfnamefont {D.~R.}\ \bibnamefont {Harshman}}, \bibinfo
  {author} {\bibfnamefont {M.}~\bibnamefont {Senba}}, \ and\ \bibinfo {author}
  {\bibfnamefont {E.~J.}\ \bibnamefont {Ansaldo}},\ }\href {\doibase
  10.1103/PhysRevB.31.546} {\bibfield  {journal} {\bibinfo  {journal} {Phys.
  Rev. B}\ }\textbf {\bibinfo {volume} {31}},\ \bibinfo {pages} {546} (\bibinfo
  {year} {1985})}\BibitemShut {NoStop}%
\bibitem [{\citenamefont {MacLaughlin}\ \emph {et~al.}(2004)\citenamefont
  {MacLaughlin}, \citenamefont {Heffner}, \citenamefont {Bernal}, \citenamefont
  {Ishida}, \citenamefont {Sonier}, \citenamefont {Nieuwenhuys}, \citenamefont
  {Maple},\ and\ \citenamefont {Stewart}}]{McLaughlin04}%
  \BibitemOpen
  \bibfield  {author} {\bibinfo {author} {\bibfnamefont {D.~E.}\ \bibnamefont
  {MacLaughlin}}, \bibinfo {author} {\bibfnamefont {R.~H.}\ \bibnamefont
  {Heffner}}, \bibinfo {author} {\bibfnamefont {O.~O.}\ \bibnamefont {Bernal}},
  \bibinfo {author} {\bibfnamefont {K.}~\bibnamefont {Ishida}}, \bibinfo
  {author} {\bibfnamefont {J.~E.}\ \bibnamefont {Sonier}}, \bibinfo {author}
  {\bibfnamefont {G.~J.}\ \bibnamefont {Nieuwenhuys}}, \bibinfo {author}
  {\bibfnamefont {M.~B.}\ \bibnamefont {Maple}}, \ and\ \bibinfo {author}
  {\bibfnamefont {G.~R.}\ \bibnamefont {Stewart}},\ }\href
  {http://stacks.iop.org/0953-8984/16/i=40/a=005} {\bibfield  {journal}
  {\bibinfo  {journal} {J. Phys. Condens. Matter}\ }\textbf {\bibinfo {volume}
  {16}},\ \bibinfo {pages} {S4479} (\bibinfo {year} {2004})}\BibitemShut
  {NoStop}%
\bibitem [{\citenamefont {Hayano}\ \emph {et~al.}(1979)\citenamefont {Hayano},
  \citenamefont {Uemura}, \citenamefont {Imazato}, \citenamefont {Nishida},
  \citenamefont {Yamazaki},\ and\ \citenamefont {Kubo}}]{Hayanoetal79}%
  \BibitemOpen
  \bibfield  {author} {\bibinfo {author} {\bibfnamefont {R.~S.}\ \bibnamefont
  {Hayano}}, \bibinfo {author} {\bibfnamefont {Y.~J.}\ \bibnamefont {Uemura}},
  \bibinfo {author} {\bibfnamefont {J.}~\bibnamefont {Imazato}}, \bibinfo
  {author} {\bibfnamefont {N.}~\bibnamefont {Nishida}}, \bibinfo {author}
  {\bibfnamefont {T.}~\bibnamefont {Yamazaki}}, \ and\ \bibinfo {author}
  {\bibfnamefont {R.}~\bibnamefont {Kubo}},\ }\href {\doibase
  10.1103/PhysRevB.20.850} {\bibfield  {journal} {\bibinfo  {journal} {Phys.
  Rev. B}\ }\textbf {\bibinfo {volume} {20}},\ \bibinfo {pages} {850} (\bibinfo
  {year} {1979})}\BibitemShut {NoStop}%
\bibitem [{\citenamefont {Noakes}\ and\ \citenamefont
  {Kalvius}(1997)}]{NoakesKalvius97}%
  \BibitemOpen
  \bibfield  {author} {\bibinfo {author} {\bibfnamefont {D.~R.}\ \bibnamefont
  {Noakes}}\ and\ \bibinfo {author} {\bibfnamefont {G.~M.}\ \bibnamefont
  {Kalvius}},\ }\href {\doibase 10.1103/PhysRevB.56.2352} {\bibfield  {journal}
  {\bibinfo  {journal} {Phys. Rev. B}\ }\textbf {\bibinfo {volume} {56}},\
  \bibinfo {pages} {2352} (\bibinfo {year} {1997})}\BibitemShut {NoStop}%
\bibitem [{Note5()}]{Note5}%
  \BibitemOpen
  \bibinfo {note} {Such a symmetry does occur in certain one-dimensional random
  spin chains \cite {Fisher92,*Fisher95}, but it does not generalize to higher
  dimensions.}\BibitemShut {Stop}%
\bibitem [{\citenamefont {Bauer}\ \emph {et~al.}(2005)\citenamefont {Bauer},
  \citenamefont {Zapf}, \citenamefont {Ho}, \citenamefont {Butch},
  \citenamefont {Freeman}, \citenamefont {Sirvent},\ and\ \citenamefont
  {Maple}}]{Baueretal05}%
  \BibitemOpen
  \bibfield  {author} {\bibinfo {author} {\bibfnamefont {E.~D.}\ \bibnamefont
  {Bauer}}, \bibinfo {author} {\bibfnamefont {V.~S.}\ \bibnamefont {Zapf}},
  \bibinfo {author} {\bibfnamefont {P.-C.}\ \bibnamefont {Ho}}, \bibinfo
  {author} {\bibfnamefont {N.~P.}\ \bibnamefont {Butch}}, \bibinfo {author}
  {\bibfnamefont {E.~J.}\ \bibnamefont {Freeman}}, \bibinfo {author}
  {\bibfnamefont {C.}~\bibnamefont {Sirvent}}, \ and\ \bibinfo {author}
  {\bibfnamefont {M.~B.}\ \bibnamefont {Maple}},\ }\href {\doibase
  10.1103/PhysRevLett.94.046401} {\bibfield  {journal} {\bibinfo  {journal}
  {Phys. Rev. Lett.}\ }\textbf {\bibinfo {volume} {94}},\ \bibinfo {pages}
  {046401} (\bibinfo {year} {2005})}\BibitemShut {NoStop}%
\bibitem [{\citenamefont {Butch}\ and\ \citenamefont
  {Maple}(2009)}]{ButchMaple09}%
  \BibitemOpen
  \bibfield  {author} {\bibinfo {author} {\bibfnamefont {N.~P.}\ \bibnamefont
  {Butch}}\ and\ \bibinfo {author} {\bibfnamefont {M.~B.}\ \bibnamefont
  {Maple}},\ }\href {\doibase 10.1103/PhysRevLett.103.076404} {\bibfield
  {journal} {\bibinfo  {journal} {Phys. Rev. Lett.}\ }\textbf {\bibinfo
  {volume} {103}},\ \bibinfo {pages} {076404} (\bibinfo {year}
  {2009})}\BibitemShut {NoStop}%
\bibitem [{\citenamefont {Gebretsadik}\ \emph {et~al.}()\citenamefont
  {Gebretsadik}, \citenamefont {Wang}, \citenamefont {Schroeder},\ and\
  \citenamefont {Page}}]{Gebretsadiketal_unpublished}%
  \BibitemOpen
  \bibfield  {author} {\bibinfo {author} {\bibfnamefont {A.}~\bibnamefont
  {Gebretsadik}}, \bibinfo {author} {\bibfnamefont {R.}~\bibnamefont {Wang}},
  \bibinfo {author} {\bibfnamefont {A.}~\bibnamefont {Schroeder}}, \ and\
  \bibinfo {author} {\bibfnamefont {K.}~\bibnamefont {Page}},\ }\href@noop {}
  {}\bibinfo {note} {To be published}\BibitemShut {NoStop}%
\bibitem [{\citenamefont {Proffen}\ \emph {et~al.}(2002)\citenamefont
  {Proffen}, \citenamefont {Egami}, \citenamefont {Billinge}, \citenamefont
  {Cheetham}, \citenamefont {Louca},\ and\ \citenamefont {Parise}}]{NPDF}%
  \BibitemOpen
  \bibfield  {author} {\bibinfo {author} {\bibfnamefont {T.}~\bibnamefont
  {Proffen}}, \bibinfo {author} {\bibfnamefont {T.}~\bibnamefont {Egami}},
  \bibinfo {author} {\bibfnamefont {S.}~\bibnamefont {Billinge}}, \bibinfo
  {author} {\bibfnamefont {A.}~\bibnamefont {Cheetham}}, \bibinfo {author}
  {\bibfnamefont {D.}~\bibnamefont {Louca}}, \ and\ \bibinfo {author}
  {\bibfnamefont {J.}~\bibnamefont {Parise}},\ }\href {\doibase
  10.1007/s003390201929} {\bibfield  {journal} {\bibinfo  {journal} {Appl.
  Phys. A}\ }\textbf {\bibinfo {volume} {74}},\ \bibinfo {pages} {s163}
  (\bibinfo {year} {2002})}\BibitemShut {NoStop}%
\bibitem [{\citenamefont {Neuefeind}\ \emph {et~al.}(2012)\citenamefont
  {Neuefeind}, \citenamefont {Feygenson}, \citenamefont {Carruth},
  \citenamefont {Hoffmann},\ and\ \citenamefont {Chipley}}]{NOMAD}%
  \BibitemOpen
  \bibfield  {author} {\bibinfo {author} {\bibfnamefont {J.}~\bibnamefont
  {Neuefeind}}, \bibinfo {author} {\bibfnamefont {M.}~\bibnamefont
  {Feygenson}}, \bibinfo {author} {\bibfnamefont {J.}~\bibnamefont {Carruth}},
  \bibinfo {author} {\bibfnamefont {R.}~\bibnamefont {Hoffmann}}, \ and\
  \bibinfo {author} {\bibfnamefont {K.~K.}\ \bibnamefont {Chipley}},\ }\href
  {\doibase http://dx.doi.org/10.1016/j.nimb.2012.05.037} {\bibfield  {journal}
  {\bibinfo  {journal} {Nuclear Instruments and Methods in Physics Research B}\
  }\textbf {\bibinfo {volume} {287}},\ \bibinfo {pages} {68 } (\bibinfo {year}
  {2012})}\BibitemShut {NoStop}%
\bibitem [{\citenamefont {Proffen}\ and\ \citenamefont
  {Billinge}(1999)}]{PDFFIT}%
  \BibitemOpen
  \bibfield  {author} {\bibinfo {author} {\bibfnamefont {T.}~\bibnamefont
  {Proffen}}\ and\ \bibinfo {author} {\bibfnamefont {S.~J.~L.}\ \bibnamefont
  {Billinge}},\ }\href {\doibase 10.1107/S0021889899003532} {\bibfield
  {journal} {\bibinfo  {journal} {J. Appl. Crystallogr.}\ }\textbf {\bibinfo
  {volume} {32}},\ \bibinfo {pages} {572} (\bibinfo {year} {1999})}\BibitemShut
  {NoStop}%
\bibitem [{\citenamefont {Farrow}\ \emph {et~al.}(2007)\citenamefont {Farrow},
  \citenamefont {Juhas}, \citenamefont {Liu}, \citenamefont {Bryndin},
  \citenamefont {Bozin}, \citenamefont {Bloch}, \citenamefont {Proffen},\ and\
  \citenamefont {Billinge}}]{PDFgui}%
  \BibitemOpen
  \bibfield  {author} {\bibinfo {author} {\bibfnamefont {C.~L.}\ \bibnamefont
  {Farrow}}, \bibinfo {author} {\bibfnamefont {P.}~\bibnamefont {Juhas}},
  \bibinfo {author} {\bibfnamefont {J.~W.}\ \bibnamefont {Liu}}, \bibinfo
  {author} {\bibfnamefont {D.}~\bibnamefont {Bryndin}}, \bibinfo {author}
  {\bibfnamefont {E.~S.}\ \bibnamefont {Bozin}}, \bibinfo {author}
  {\bibfnamefont {J.}~\bibnamefont {Bloch}}, \bibinfo {author} {\bibfnamefont
  {T.}~\bibnamefont {Proffen}}, \ and\ \bibinfo {author} {\bibfnamefont
  {S.~J.~L.}\ \bibnamefont {Billinge}},\ }\href
  {http://stacks.iop.org/0953-8984/19/i=33/a=335219} {\bibfield  {journal}
  {\bibinfo  {journal} {J. Phys. Condens. Matter}\ }\textbf {\bibinfo {volume}
  {19}},\ \bibinfo {pages} {335219} (\bibinfo {year} {2007})}\BibitemShut
  {NoStop}%
\bibitem [{\citenamefont {Proffen}\ \emph {et~al.}(2009)\citenamefont
  {Proffen}, \citenamefont {Billinge},\ and\ \citenamefont {Vogt}}]{Cu3Au}%
  \BibitemOpen
  \bibfield  {author} {\bibinfo {author} {\bibfnamefont {T.}~\bibnamefont
  {Proffen}}, \bibinfo {author} {\bibfnamefont {V.}~\bibnamefont
  {Petkov}}, \bibinfo {author} {\bibfnamefont {S.~J.~L.}\ \bibnamefont
  {Billinge}}, \ and\ \bibinfo {author} {\bibfnamefont {T.}~\bibnamefont
  {Vogt}},\ }\href {\doibase 10.1524/zkri.217.2.47.20626} {\bibfield  {journal}
  {\bibinfo  {journal} {Zeitschrift f\"ur Kristallographie - Crystalline
  Materials}\ }\textbf {\bibinfo {volume} {217}},\ \bibinfo {pages} {47}
  (\bibinfo {year} {2009})}\BibitemShut {NoStop}%
\bibitem [{\citenamefont {Liu}\ \emph {et~al.}(2016)\citenamefont {Liu},
  \citenamefont {Huq}, \citenamefont {Moorhead-Rosenberg}, \citenamefont
  {Manthiram},\ and\ \citenamefont {Page}}]{LMNO}%
  \BibitemOpen
  \bibfield  {author} {\bibinfo {author} {\bibfnamefont {J.}~\bibnamefont
  {Liu}}, \bibinfo {author} {\bibfnamefont {A.}~\bibnamefont {Huq}}, \bibinfo
  {author} {\bibfnamefont {Z.}~\bibnamefont {Moorhead-Rosenberg}}, \bibinfo
  {author} {\bibfnamefont {A.}~\bibnamefont {Manthiram}}, \ and\ \bibinfo
  {author} {\bibfnamefont {K.}~\bibnamefont {Page}},\ }\href {\doibase
  10.1021/acs.chemmater.6b02946} {\bibfield  {journal} {\bibinfo  {journal}
  {Chemistry of Materials}\ }\textbf {\bibinfo {volume} {28}},\ \bibinfo
  {pages} {6817} (\bibinfo {year} {2016})}\BibitemShut {NoStop}%
\bibitem [{\citenamefont {Kornilov}\ and\ \citenamefont
  {Pomjakushin}(1991)}]{Kornilov}%
  \BibitemOpen
  \bibfield  {author} {\bibinfo {author} {\bibfnamefont {E.}~\bibnamefont
  {Kornilov}}\ and\ \bibinfo {author} {\bibfnamefont {V.}~\bibnamefont
  {Pomjakushin}},\ }\href {\doibase
  http://dx.doi.org/10.1016/0375-9601(91)90959-C} {\bibfield  {journal}
  {\bibinfo  {journal} {Physics Letters A}\ }\textbf {\bibinfo {volume}
  {153}},\ \bibinfo {pages} {364 } (\bibinfo {year} {1991})}\BibitemShut
  {NoStop}%
\bibitem [{\citenamefont {Pratt}(2000)}]{Pratt2000}%
  \BibitemOpen
  \bibfield  {author} {\bibinfo {author} {\bibfnamefont {F.}~\bibnamefont
  {Pratt}},\ }\href {\doibase http://dx.doi.org/10.1016/S0921-4526(00)00328-8}
  {\bibfield  {journal} {\bibinfo  {journal} {Physica B}\ }\textbf {\bibinfo
  {volume} {289-290}},\ \bibinfo {pages} {710 } (\bibinfo {year}
  {2000})}\BibitemShut {NoStop}%
\bibitem [{\citenamefont {Nozadze}\ and\ \citenamefont
  {Vojta}(2012)}]{NozadzeVojta12}%
  \BibitemOpen
  \bibfield  {author} {\bibinfo {author} {\bibfnamefont {D.}~\bibnamefont
  {Nozadze}}\ and\ \bibinfo {author} {\bibfnamefont {T.}~\bibnamefont
  {Vojta}},\ }\href {\doibase 10.1103/PhysRevB.85.174202} {\bibfield  {journal}
  {\bibinfo  {journal} {Phys. Rev. B}\ }\textbf {\bibinfo {volume} {85}},\
  \bibinfo {pages} {174202} (\bibinfo {year} {2012})}\BibitemShut {NoStop}%
\end{thebibliography}

\begin{thebibliography}{17}%
\makeatletter
\providecommand \@ifxundefined [1]{%
 \@ifx{#1\undefined}
}%
\providecommand \@ifnum [1]{%
 \ifnum #1\expandafter \@firstoftwo
 \else \expandafter \@secondoftwo
 \fi
}%
\providecommand \@ifx [1]{%
 \ifx #1\expandafter \@firstoftwo
 \else \expandafter \@secondoftwo
 \fi
}%
\providecommand \natexlab [1]{#1}%
\providecommand \enquote  [1]{``#1''}%
\providecommand \bibnamefont  [1]{#1}%
\providecommand \bibfnamefont [1]{#1}%
\providecommand \citenamefont [1]{#1}%
\providecommand \href@noop [0]{\@secondoftwo}%
\providecommand \href [0]{\begingroup \@sanitize@url \@href}%
\providecommand \@href[1]{\@@startlink{#1}\@@href}%
\providecommand \@@href[1]{\endgroup#1\@@endlink}%
\providecommand \@sanitize@url [0]{\catcode `\\12\catcode `\$12\catcode
  `\&12\catcode `\#12\catcode `\^12\catcode `\_12\catcode `\%12\relax}%
\providecommand \@@startlink[1]{}%
\providecommand \@@endlink[0]{}%
\providecommand \url  [0]{\begingroup\@sanitize@url \@url }%
\providecommand \@url [1]{\endgroup\@href {#1}{\urlprefix }}%
\providecommand \urlprefix  [0]{URL }%
\providecommand \Eprint [0]{\href }%
\providecommand \doibase [0]{http://dx.doi.org/}%
\providecommand \selectlanguage [0]{\@gobble}%
\providecommand \bibinfo  [0]{\@secondoftwo}%
\providecommand \bibfield  [0]{\@secondoftwo}%
\providecommand \translation [1]{[#1]}%
\providecommand \BibitemOpen [0]{}%
\providecommand \bibitemStop [0]{}%
\providecommand \bibitemNoStop [0]{.\EOS\space}%
\providecommand \EOS [0]{\spacefactor3000\relax}%
\providecommand \BibitemShut  [1]{\csname bibitem#1\endcsname}%
\let\auto@bib@innerbib\@empty
%</preamble>
\bibitem [{\citenamefont {Ubaid-Kassis}\ \emph {et~al.}(2010)\citenamefont
  {Ubaid-Kassis}, \citenamefont {Vojta},\ and\ \citenamefont
  {Schroeder}}]{UbaidKassisVojtaSchroeder10sup}%
  \BibitemOpen
  \bibfield  {author} {\bibinfo {author} {\bibfnamefont {S.}~\bibnamefont
  {Ubaid-Kassis}}, \bibinfo {author} {\bibfnamefont {T.}~\bibnamefont {Vojta}},
  \ and\ \bibinfo {author} {\bibfnamefont {A.}~\bibnamefont {Schroeder}},\
  }\href@noop {} {\bibfield  {journal} {\bibinfo  {journal} {Phys. Rev. Lett.}\
  }\textbf {\bibinfo {volume} {104}},\ \bibinfo {pages} {066402} (\bibinfo
  {year} {2010})}\BibitemShut {NoStop}%
\bibitem [{\citenamefont {Schroeder}\ \emph {et~al.}(2014)\citenamefont
  {Schroeder}, \citenamefont {Wang}, \citenamefont {Baker}, \citenamefont
  {Pratt}, \citenamefont {Blundell}, \citenamefont {Lancaster}, \citenamefont
  {Franke},\ and\ \citenamefont {M\"oller}}]{Schroederetal14sup}%
  \BibitemOpen
  \bibfield  {author} {\bibinfo {author} {\bibfnamefont {A.}~\bibnamefont
  {Schroeder}}, \bibinfo {author} {\bibfnamefont {R.}~\bibnamefont {Wang}},
  \bibinfo {author} {\bibfnamefont {P.~J.}\ \bibnamefont {Baker}}, \bibinfo
  {author} {\bibfnamefont {F.~L.}\ \bibnamefont {Pratt}}, \bibinfo {author}
  {\bibfnamefont {S.~J.}\ \bibnamefont {Blundell}}, \bibinfo {author}
  {\bibfnamefont {T.}~\bibnamefont {Lancaster}}, \bibinfo {author}
  {\bibfnamefont {I.}~\bibnamefont {Franke}}, \ and\ \bibinfo {author}
  {\bibfnamefont {J.~S.}\ \bibnamefont {M\"oller}},\ }\href
  {http://stacks.iop.org/1742-6596/551/i=1/a=012003} {\bibfield  {journal}
  {\bibinfo  {journal} {J. Phys. Conf. Series}\ }\textbf {\bibinfo {volume}
  {551}},\ \bibinfo {pages} {012003} (\bibinfo {year} {2014})}\BibitemShut
  {NoStop}%
\bibitem [{\citenamefont {Gebretsadik}\ \emph {et~al.}()\citenamefont
  {Gebretsadik}, \citenamefont {Wang}, \citenamefont {Schroeder},\ and\
  \citenamefont {Page}}]{Gebretsadiketal_unpublishedsup}%
  \BibitemOpen
  \bibfield  {author} {\bibinfo {author} {\bibfnamefont {A.}~\bibnamefont
  {Gebretsadik}}, \bibinfo {author} {\bibfnamefont {R.}~\bibnamefont {Wang}},
  \bibinfo {author} {\bibfnamefont {A.}~\bibnamefont {Schroeder}}, \ and\
  \bibinfo {author} {\bibfnamefont {K.}~\bibnamefont {Page}},\ }\href@noop {}
  {}\bibinfo {note} {To be published}\BibitemShut {NoStop}%
\bibitem [{\citenamefont {Proffen}\ \emph {et~al.}(2002)\citenamefont
  {Proffen}, \citenamefont {Egami}, \citenamefont {Billinge}, \citenamefont
  {Cheetham}, \citenamefont {Louca},\ and\ \citenamefont {Parise}}]{NPDFsup}%
  \BibitemOpen
  \bibfield  {author} {\bibinfo {author} {\bibfnamefont {T.}~\bibnamefont
  {Proffen}}, \bibinfo {author} {\bibfnamefont {T.}~\bibnamefont {Egami}},
  \bibinfo {author} {\bibfnamefont {S.}~\bibnamefont {Billinge}}, \bibinfo
  {author} {\bibfnamefont {A.}~\bibnamefont {Cheetham}}, \bibinfo {author}
  {\bibfnamefont {D.}~\bibnamefont {Louca}}, \ and\ \bibinfo {author}
  {\bibfnamefont {J.}~\bibnamefont {Parise}},\ }\href {\doibase
  10.1007/s003390201929} {\bibfield  {journal} {\bibinfo  {journal} {Appl.
  Phys. A}\ }\textbf {\bibinfo {volume} {74}},\ \bibinfo {pages} {s163}
  (\bibinfo {year} {2002})}\BibitemShut {NoStop}%
\bibitem [{\citenamefont {Neuefeind}\ \emph {et~al.}(2012)\citenamefont
  {Neuefeind}, \citenamefont {Feygenson}, \citenamefont {Carruth},
  \citenamefont {Hoffmann},\ and\ \citenamefont {Chipley}}]{NOMADsup}%
  \BibitemOpen
  \bibfield  {author} {\bibinfo {author} {\bibfnamefont {J.}~\bibnamefont
  {Neuefeind}}, \bibinfo {author} {\bibfnamefont {M.}~\bibnamefont
  {Feygenson}}, \bibinfo {author} {\bibfnamefont {J.}~\bibnamefont {Carruth}},
  \bibinfo {author} {\bibfnamefont {R.}~\bibnamefont {Hoffmann}}, \ and\
  \bibinfo {author} {\bibfnamefont {K.~K.}\ \bibnamefont {Chipley}},\ }\href
  {\doibase http://dx.doi.org/10.1016/j.nimb.2012.05.037} {\bibfield  {journal}
  {\bibinfo  {journal} {Nuclear Instruments and Methods in Physics Research B}\
  }\textbf {\bibinfo {volume} {287}},\ \bibinfo {pages} {68 } (\bibinfo {year}
  {2012})}\BibitemShut {NoStop}%
\bibitem [{\citenamefont {Proffen}\ and\ \citenamefont
  {Billinge}(1999)}]{PDFFITsup}%
  \BibitemOpen
  \bibfield  {author} {\bibinfo {author} {\bibfnamefont {T.}~\bibnamefont
  {Proffen}}\ and\ \bibinfo {author} {\bibfnamefont {S.~J.~L.}\ \bibnamefont
  {Billinge}},\ }\href {\doibase 10.1107/S0021889899003532} {\bibfield
  {journal} {\bibinfo  {journal} {J. Appl. Crystallogr.}\ }\textbf {\bibinfo
  {volume} {32}},\ \bibinfo {pages} {572} (\bibinfo {year} {1999})}\BibitemShut
  {NoStop}%
\bibitem [{\citenamefont {Farrow}\ \emph {et~al.}(2007)\citenamefont {Farrow},
  \citenamefont {Juhas}, \citenamefont {Liu}, \citenamefont {Bryndin},
  \citenamefont {Bozin}, \citenamefont {Bloch}, \citenamefont {Proffen},\ and\
  \citenamefont {Billinge}}]{PDFguisup}%
  \BibitemOpen
  \bibfield  {author} {\bibinfo {author} {\bibfnamefont {C.~L.}\ \bibnamefont
  {Farrow}}, \bibinfo {author} {\bibfnamefont {P.}~\bibnamefont {Juhas}},
  \bibinfo {author} {\bibfnamefont {J.~W.}\ \bibnamefont {Liu}}, \bibinfo
  {author} {\bibfnamefont {D.}~\bibnamefont {Bryndin}}, \bibinfo {author}
  {\bibfnamefont {E.~S.}\ \bibnamefont {Bozin}}, \bibinfo {author}
  {\bibfnamefont {J.}~\bibnamefont {Bloch}}, \bibinfo {author} {\bibfnamefont
  {T.}~\bibnamefont {Proffen}}, \ and\ \bibinfo {author} {\bibfnamefont
  {S.~J.~L.}\ \bibnamefont {Billinge}},\ }\href
  {http://stacks.iop.org/0953-8984/19/i=33/a=335219} {\bibfield  {journal}
  {\bibinfo  {journal} {J. Phys. Condens. Matter}\ }\textbf {\bibinfo {volume}
  {19}},\ \bibinfo {pages} {335219} (\bibinfo {year} {2007})}\BibitemShut
  {NoStop}%
\bibitem [{\citenamefont {Proffen}\ \emph {et~al.}(2009)\citenamefont
  {Proffen}, \citenamefont {Billinge},\ and\ \citenamefont {Vogt}}]{Cu3Ausup}%
  \BibitemOpen
  \bibfield  {author} {\bibinfo {author} {\bibfnamefont {T.}~\bibnamefont
  {Proffen}}, \bibinfo {author} {\bibfnamefont {V.}~\bibnamefont
  {Petkov}}, \bibinfo {author} {\bibfnamefont {S.~J.~L.}\ \bibnamefont
  {Billinge}}, \ and\ \bibinfo {author} {\bibfnamefont {T.}~\bibnamefont
  {Vogt}},\ }\href {\doibase 10.1524/zkri.217.2.47.20626} {\bibfield  {journal}
  {\bibinfo  {journal} {Zeitschrift f\"ur Kristallographie - Crystalline
  Materials}\ }\textbf {\bibinfo {volume} {217}},\ \bibinfo {pages} {47}
  (\bibinfo {year} {2009})}\BibitemShut {NoStop}%
\bibitem [{\citenamefont {Liu}\ \emph {et~al.}(2016)\citenamefont {Liu},
  \citenamefont {Huq}, \citenamefont {Moorhead-Rosenberg}, \citenamefont
  {Manthiram},\ and\ \citenamefont {Page}}]{LMNOsup}%
  \BibitemOpen
  \bibfield  {author} {\bibinfo {author} {\bibfnamefont {J.}~\bibnamefont
  {Liu}}, \bibinfo {author} {\bibfnamefont {A.}~\bibnamefont {Huq}}, \bibinfo
  {author} {\bibfnamefont {Z.}~\bibnamefont {Moorhead-Rosenberg}}, \bibinfo
  {author} {\bibfnamefont {A.}~\bibnamefont {Manthiram}}, \ and\ \bibinfo
  {author} {\bibfnamefont {K.}~\bibnamefont {Page}},\ }\href {\doibase
  10.1021/acs.chemmater.6b02946} {\bibfield  {journal} {\bibinfo  {journal}
  {Chemistry of Materials}\ }\textbf {\bibinfo {volume} {28}},\ \bibinfo
  {pages} {6817} (\bibinfo {year} {2016})}\BibitemShut {NoStop}%
\bibitem [{\citenamefont {Hayano}\ \emph {et~al.}(1979)\citenamefont {Hayano},
  \citenamefont {Uemura}, \citenamefont {Imazato}, \citenamefont {Nishida},
  \citenamefont {Yamazaki},\ and\ \citenamefont {Kubo}}]{Hayanoetal79sup}%
  \BibitemOpen
  \bibfield  {author} {\bibinfo {author} {\bibfnamefont {R.~S.}\ \bibnamefont
  {Hayano}}, \bibinfo {author} {\bibfnamefont {Y.~J.}\ \bibnamefont {Uemura}},
  \bibinfo {author} {\bibfnamefont {J.}~\bibnamefont {Imazato}}, \bibinfo
  {author} {\bibfnamefont {N.}~\bibnamefont {Nishida}}, \bibinfo {author}
  {\bibfnamefont {T.}~\bibnamefont {Yamazaki}}, \ and\ \bibinfo {author}
  {\bibfnamefont {R.}~\bibnamefont {Kubo}},\ }\href {\doibase
  10.1103/PhysRevB.20.850} {\bibfield  {journal} {\bibinfo  {journal} {Phys.
  Rev. B}\ }\textbf {\bibinfo {volume} {20}},\ \bibinfo {pages} {850} (\bibinfo
  {year} {1979})}\BibitemShut {NoStop}%
\bibitem [{\citenamefont {Kornilov}\ and\ \citenamefont
  {Pomjakushin}(1991)}]{Kornilovsup}%
  \BibitemOpen
  \bibfield  {author} {\bibinfo {author} {\bibfnamefont {E.}~\bibnamefont
  {Kornilov}}\ and\ \bibinfo {author} {\bibfnamefont {V.}~\bibnamefont
  {Pomjakushin}},\ }\href {\doibase
  http://dx.doi.org/10.1016/0375-9601(91)90959-C} {\bibfield  {journal}
  {\bibinfo  {journal} {Physics Letters A}\ }\textbf {\bibinfo {volume}
  {153}},\ \bibinfo {pages} {364 } (\bibinfo {year} {1991})}\BibitemShut
  {NoStop}%
\bibitem [{\citenamefont {Yaouanc}\ and\ \citenamefont {Dalmas~de
  Reotier}(2011)}]{YaouancReotier11sup}%
  \BibitemOpen
  \bibfield  {author} {\bibinfo {author} {\bibfnamefont {A.}~\bibnamefont
  {Yaouanc}}\ and\ \bibinfo {author} {\bibfnamefont {P.}~\bibnamefont
  {Dalmas~de Reotier}},\ }\href@noop {} {\emph {\bibinfo {title} {Muon Spin
  Rotation, Relaxation, and Resonance: Applications to Condensed Matter}}}\
  (\bibinfo  {publisher} {Oxford University Press},\ \bibinfo {address}
  {Oxford},\ \bibinfo {year} {2011})\BibitemShut {NoStop}%
\bibitem [{\citenamefont {Pratt}(2000)}]{Pratt2000sup}%
  \BibitemOpen
  \bibfield  {author} {\bibinfo {author} {\bibfnamefont {F.}~\bibnamefont
  {Pratt}},\ }\href {\doibase http://dx.doi.org/10.1016/S0921-4526(00)00328-8}
  {\bibfield  {journal} {\bibinfo  {journal} {Physica B}\ }\textbf {\bibinfo
  {volume} {289-290}},\ \bibinfo {pages} {710 } (\bibinfo {year}
  {2000})}\BibitemShut {NoStop}%
\bibitem [{\citenamefont {Noakes}\ and\ \citenamefont
  {Kalvius}(1997)}]{NoakesKalvius97sup}%
  \BibitemOpen
  \bibfield  {author} {\bibinfo {author} {\bibfnamefont {D.~R.}\ \bibnamefont
  {Noakes}}\ and\ \bibinfo {author} {\bibfnamefont {G.~M.}\ \bibnamefont
  {Kalvius}},\ }\href {\doibase 10.1103/PhysRevB.56.2352} {\bibfield  {journal}
  {\bibinfo  {journal} {Phys. Rev. B}\ }\textbf {\bibinfo {volume} {56}},\
  \bibinfo {pages} {2352} (\bibinfo {year} {1997})}\BibitemShut {NoStop}%
\bibitem [{\citenamefont {Wang}\ \emph {et~al.}(2015)\citenamefont {Wang},
  \citenamefont {Ubaid-Kassis}, \citenamefont {Schroeder}, \citenamefont
  {Baker}, \citenamefont {Pratt}, \citenamefont {Blundell}, \citenamefont
  {Lancaster}, \citenamefont {Franke}, \citenamefont {M\"oller},\ and\
  \citenamefont {Vojta}}]{Wangetal15sup}%
  \BibitemOpen
  \bibfield  {author} {\bibinfo {author} {\bibfnamefont {R.}~\bibnamefont
  {Wang}}, \bibinfo {author} {\bibfnamefont {S.}~\bibnamefont {Ubaid-Kassis}},
  \bibinfo {author} {\bibfnamefont {A.}~\bibnamefont {Schroeder}}, \bibinfo
  {author} {\bibfnamefont {P.}~\bibnamefont {Baker}}, \bibinfo {author}
  {\bibfnamefont {F.}~\bibnamefont {Pratt}}, \bibinfo {author} {\bibfnamefont
  {S.}~\bibnamefont {Blundell}}, \bibinfo {author} {\bibfnamefont
  {T.}~\bibnamefont {Lancaster}}, \bibinfo {author} {\bibfnamefont
  {I.}~\bibnamefont {Franke}}, \bibinfo {author} {\bibfnamefont
  {J.}~\bibnamefont {M\"oller}}, \ and\ \bibinfo {author} {\bibfnamefont
  {T.}~\bibnamefont {Vojta}},\ }\href
  {http://stacks.iop.org/1742-6596/592/i=1/a=012089} {\bibfield  {journal}
  {\bibinfo  {journal} {J. Phys. Conf. Series}\ }\textbf {\bibinfo {volume}
  {592}},\ \bibinfo {pages} {012089} (\bibinfo {year} {2015})}\BibitemShut
  {NoStop}%
\bibitem [{\citenamefont {Uemura}\ \emph {et~al.}(1985)\citenamefont {Uemura},
  \citenamefont {Yamazaki}, \citenamefont {Harshman}, \citenamefont {Senba},\
  and\ \citenamefont {Ansaldo}}]{uemuraSGsup}%
  \BibitemOpen
  \bibfield  {author} {\bibinfo {author} {\bibfnamefont {Y.~J.}\ \bibnamefont
  {Uemura}}, \bibinfo {author} {\bibfnamefont {T.}~\bibnamefont {Yamazaki}},
  \bibinfo {author} {\bibfnamefont {D.~R.}\ \bibnamefont {Harshman}}, \bibinfo
  {author} {\bibfnamefont {M.}~\bibnamefont {Senba}}, \ and\ \bibinfo {author}
  {\bibfnamefont {E.~J.}\ \bibnamefont {Ansaldo}},\ }\href {\doibase
  10.1103/PhysRevB.31.546} {\bibfield  {journal} {\bibinfo  {journal} {Phys.
  Rev. B}\ }\textbf {\bibinfo {volume} {31}},\ \bibinfo {pages} {546} (\bibinfo
  {year} {1985})}\BibitemShut {NoStop}%
\bibitem [{\citenamefont {MacLaughlin}\ \emph {et~al.}(2004)\citenamefont
  {MacLaughlin}, \citenamefont {Heffner}, \citenamefont {Bernal}, \citenamefont
  {Ishida}, \citenamefont {Sonier}, \citenamefont {Nieuwenhuys}, \citenamefont
  {Maple},\ and\ \citenamefont {Stewart}}]{McLaughlin04sup}%
  \BibitemOpen
  \bibfield  {author} {\bibinfo {author} {\bibfnamefont {D.~E.}\ \bibnamefont
  {MacLaughlin}}, \bibinfo {author} {\bibfnamefont {R.~H.}\ \bibnamefont
  {Heffner}}, \bibinfo {author} {\bibfnamefont {O.~O.}\ \bibnamefont {Bernal}},
  \bibinfo {author} {\bibfnamefont {K.}~\bibnamefont {Ishida}}, \bibinfo
  {author} {\bibfnamefont {J.~E.}\ \bibnamefont {Sonier}}, \bibinfo {author}
  {\bibfnamefont {G.~J.}\ \bibnamefont {Nieuwenhuys}}, \bibinfo {author}
  {\bibfnamefont {M.~B.}\ \bibnamefont {Maple}}, \ and\ \bibinfo {author}
  {\bibfnamefont {G.~R.}\ \bibnamefont {Stewart}},\ }\href
  {http://stacks.iop.org/0953-8984/16/i=40/a=005} {\bibfield  {journal}
  {\bibinfo  {journal} {J. Phys. Condens. Matter}\ }\textbf {\bibinfo {volume}
  {16}},\ \bibinfo {pages} {S4479} (\bibinfo {year} {2004})}\BibitemShut
  {NoStop}%
\end{thebibliography}
%

\end{document}